\documentclass{aa}  

\usepackage{xcolor}
\usepackage{graphicx}
\usepackage{textcomp}
\usepackage{amssymb}
\usepackage{ulem}

\usepackage{multicol}
\usepackage{mathptmx}
\usepackage{subfig}

\newcommand{\comment}[1]{}
\usepackage{hyperref}
\def\Msun{\>{\rm M_{\odot}}}

\begin{document}

   \title{Spatial field reconstruction with INLA:}

   \subtitle{Application to simulated galaxies}

   \author{Majda Smole
          \inst{1}\thanks{email: msmole@aob.rs}, 
          Jo\~ao Rino-Silvestre\inst{2},
          Santiago Gonz\'alez-Gait\'an\inst{2}
          \and
          Marko Stalevski\inst{1,3}
          }

   \institute{Astronomical Observatory,
              Volgina 7, 11060 Belgrade, Serbia
                \and
             CENTRA, Instituto Superior T\'ecnico, Av. Rovisco Pais, Lisboa, 1049-001, Portugal
             \and
             Sterrenkundig Observatorium, Universiteit Gent, Krijgslaan 281-S9, Gent, 9000, Belgium
             }

   \date{\today}

 
  \abstract
   {}
   {Monte Carlo Radiative Transfer (MCRT) simulations are a powerful tool for understanding the role of dust in astrophysical systems and its influence on observations. However, due to the strong coupling of the radiation field and medium across the whole computational domain, the problem is non-local and non-linear and such simulations are computationally expensive in case of realistic 3D inhomogeneous dust distributions.
   We explore a novel technique for post-processing MCRT output to reduce the total computational run time by enhancing the output of computationally less expensive simulations of lower-quality. 
   }
   {   We combine principal
component analysis (PCA) and non-negative matrix factorization (NMF) as dimensionality reduction techniques together
with Gaussian Markov random fields and the Integrated nested Laplace approximation (INLA), an approximate method for Bayesian inference, to detect and reconstruct the non-random spatial structure in the images of lower signal-to-noise or with missing data.
   }
   {We test our methodology using synthetic observations of a galaxy from the SKIRT Auriga project - a suite of high resolution magneto-hydrodynamic Milky Way-sized galaxies simulated in cosmological environment by 'zoom-in' technique. With this approach, we are able to 
   reproduce high photon number reference images $\sim5$ times faster with median residuals below $\sim20\%$.

   }
   {}

   \keywords{Galaxies: general, dust, extinction, radiative transfer, Methods: numerical, Techniques: image processing}

   \maketitle
%

\section{Introduction}

Cosmological simulations provide a powerful tool for understanding galaxy formation, their evolution, and cosmic structure
on large scales. State-of-the-art hydrodynamical simulations use
subgrid physics for processes as radiative cooling, star formation, metal enrichment, and active galactic nuclei (AGN) feed-
back. Examples
of such hydrodynamical cosmological simulations include 
Illustris
(\citealt{illustris1,illustris2}; \citealt{Genel}),
EAGLE (\citealt{eagle1}; \citealt{eagle2})
and  IllustrisTNG projects (\citealt{illustris-tng1}; \citealt{illustris-tng2}).

Both observational and simulated data can be fully 
exploited only by their detailed comparison. 
Observations are required to test the results of numerical simulations. On the other 
hand, simulations can support the interpretation
of observations and constrain predictions for 
the future instruments and observational missions.
One of the challenges includes the possibility
that the observed radiation may have been altered by the medium between the emitting source and the observer.
Given the high abundance of dust in the interstellar medium, dust grains have considerable importance in shaping the observed signal, absorbing and scattering ultraviolet and optical radiation, and then re-emitting the absorbed energy at infrared wavelengths.
Thus, the observed signal provides information about the emitting source and at the same time reveals the properties of the medium on its path.

Synthetic observations can be used to fill a gap between properties derived from numerical simulations and real observations. 
They are generated, for example, starting from a snapshot of an output of a hydrodynamical simulation and solving the radiative transfer (RT) problem by simulating how radiation propagates from an astrophysical source to the observer, optionally taking into account the degradation of the image due to transfer through the atmosphere and optics of the instrument.
RT simulations play an important role in understanding the effects of dust on astrophysical observations. However, realistic dusty media are often complex and inhomogeneous, and calculating the propagation of radiation in such systems requires the implementation of special numerical methods, such as Monte Carlo radiative transfer (MCRT).
These simulations use a large number of photon packages and emulate the relevant physical processes to mimic the propagation of real photons in dusty environments.
We refer to \cite{Steinacker} for a review of numerical solution techniques for 3D dust RT problems, and \cite{Noebauer} for a recent review of MCRT codes.

Different studies offer publicly available synthetic observational data for cosmological  simulations, using the zoom-in technique. 
This technique involves a process when an additional 
zoom-in simulation with higher mass and/or time resolution is performed for a selected part of the original simulation.
Initial conditions are taken directly from a cosmological
simulation, with further implementation of subgrid physics of relevant
processes that cannot be resolved directly in the simulations.
Recent examples include studies by \cite{Kapoor} and \cite{camps2022}
that provide high resolution broadband images
of galaxies in the Auriga (\citealt{auriga-project}) and ARTEMIS 
(\citealt{artemis}) projects, respectively.

However, even with state-of-the-art MCRT codes, simulating high resolution synthetic images proves to be a computationally challenging task (\citealt{Steinacker}).
Memory requirements and simulation running times increase with the volume grid density and with the amount of photon packages used. 
In this work our goal is to develop a MCRT post-processing
methodology  able to achieve the quality of high photon number 
images, using low photon number images as an input. 
We explore the potential of Integrated nested Laplace approximation (INLA,
\citealt{inla,r-inla}), an approximate method for Bayesian inference, coupled to Gaussian Markov Random Fields
to detect and reconstruct the non-random spatial structure of simulated
galaxies.
The capability of this INLA method to reveal the underlying spatial 
structure of astronomical data has already been tested (\citealt{santiago}). Applied to integral field units (IFU)
galaxy data, INLA is able to recover structures otherwise hidden, even with highly sparse spatial information. 
The INLA method has been successfully applied to various research fields,
few examples include identifying the climatic drivers of honey bee disease in England and Wales (\citealt{bees}), forecasting deforestation in the Brazilian Amazon (\citealt{amazon}) and  mapping air pollution (\citealt{air}).

Alongside the work presented here, we developed an alternative post-processing methodology for MCRT codes, combining an auto-encoder neural network and INLA \citep{RinoSilvestre}.
Here we offer another methodology combining  principal
component analysis (PCA)  and non-negative matrix factorization (NMF) as dimensionality reduction techniques together with INLA, and test the performance using high resolution broadband images of galaxies from the Auriga project (\citealt{auriga-project}).

In section \ref{method} we introduce the employed 
techniques and codes, together with datasets used in this work,
and describe their implementation
in our methodology.
We apply our method to a galaxy from the Auriga project.
Results are presented and discussed in section \ref{results}.
In section \ref{conclusions} we summarise the main findings
of this work and draw conclusions.

\section{Methods}
\label{method}

\subsection{Simulated Data}\label{data}
\subsubsection{\textsc{skirt} code}
\label{skirt}

The \textsc{skirt}\footnote{\url{http://www.skirt.ugent.be}} is a flexible MCRT code which can simulate the propagation of the radiation through dusty media with arbitrary 3D geometries
(\citealt{Baes-Camps2015}; \citealt{Camps-Baes2015}).
Trajectories of the photon packages, emitted by primary sources, are calculated by taking into account scattering, absorption and thermal re-emission by dust grains. Their progress through a dusty medium is calculated statistically, generating random numbers from the appropriate probability density function for each of the relevant processes. Since the interaction of each individual photon package with the medium is obtained using probability density functions, the number of photon packages in the simulation has to be large enough to provide an accurate description of the radiation field.

The \textsc{skirt} code has been widely used 
in studying dust properties in spiral galaxies
(\citealt{baes2010}; \citealt{delooze2012b,delooze2012a}; \citealt{geyter}), sub-millimetre galaxies (\citealt{alma}),
dust-driven properties of high redshift galaxies (\citealt{Behrens} ;
(\citealt{Vijayan}),
galaxy evolution (\citealt{Whitney}; \citealt{Deeley}; \citealt{Zanella}), dust attenuation law in 
high redshift quasars (\citealt{DiMascia})
and investigating AGN structure (\citealt{stalevski2016};
\citealt{Stalevski2017}; \citealt{Stalevski2019}), to name a few applications\footnote{For a full list, see \url{https://skirt.ugent.be/root/_publications_all_date.html}}.

\subsubsection{Auriga galaxies}
\label{au-project}

The Auriga project\footnote{\url{https://www.mpa-garching.mpg.de/auriga/}}
provides a set of cosmological magneto-hydrodynamical zoom-in  simulations of Milky Way (MW) type galaxies (\citealt{auriga-project}). Simulations were performed using the hydrodynamic moving mesh code \textsc{arepo} (\citealt{springel2010}), with a complete treatment of galaxy formation and evolution process such as gas cooling and heating, star formation, black hole and stellar feedback, etc (\citealt{Vogelsberger}; \citealt{Genel}).

The \textsc{skirt} Auriga project \footnote{\url{https://www.auriga.ugent.be/Home--SKIRT-Auriga-Project.html}} offers high resolution synthetic images of 30 simulated  MW type galaxies from the Auriga project (\citealt{Kapoor}). This synthetic observational data, obtained with \textsc{skirt}, represents a sample of $z=0$ galaxies in the broad wavelength range from UV to sub-mm.
For each galaxy, sets of star particles and gas cells are extracted from the Auriga simulation snapshots. Star particles with an age above 10 Myr are assigned a 
spectral energy distribution (SED) using \cite{Bruzual} stellar template library,
based on metallicity and age.
Star particles with an age below 10 Myr are assumed to be 
surrounded by dust in star-forming regions, and they are assigned SED from the MAPPINGS III (\citealt{Groves}) template library.
The initial dust distribution is calculated based on the 
metallicity and gas density extracted from the Auriga snapshots, using 
the THEMIS dust model (\citealt{Jones}).
We refer to \cite{Kapoor} for detailed recipes for 
the transformation from the set of particles extracted from Auriga snapshot to the set of particles used in \textsc{skirt} simulation.

Synthetic images from the \textsc{skirt} Auriga project are publicly available in the flexible image transport system (FITS) format
composed of 2D spatial distributions at 50 different
wavelength bins. Three different inclinations of the galaxy i.e. the angle between the angular momentum
vector and the direction towards the observer comprise the dataset of each galaxy:  'face-on' ($i=0^{\circ}$), 'edge-on' ($i=90^{\circ}$) and 'intermediate' ($i=115^{\circ}$).
'Intermediate' viewing angle corresponds to the original orientation of the galaxy in the simulation box.

\subsection{INLA}

The Integrated nested Laplace approximations  (INLA; \citealt{inla}) 
is a computational method for approximate Bayesian inference of Latent Gaussian fields.
Bayesian inference is a process of deducing a probability distribution 
from data using Bayes theorem. Bayesian inference (equation \ref{Bayes}) calculates a posterior distribution $\pi(\theta~ |~ y)$,
for the given model likelihood $\pi(y ~| ~\theta)$, 
prior distribution $\pi(\theta)$ and marginal likelihood
$\pi(y)=\int{\pi(y ~|~ \theta)\pi(\theta)\textrm{d}\theta}$

\begin{align}
  \pi(\theta ~|~ y) = \frac{\pi(y ~|~ \theta)}{\pi(y)}\pi(\theta).
  \label{Bayes}
\end{align}

Most techniques for calculating posterior distribution
rely on Markov chain Monte Carlo (MCMC) methods (\citealt{MCMC}). In this class of sampling-based numerical methods, the posterior distribution is obtained after many iterations, which is often computationally expensive.
INLA provides a novel approach for faster Bayesian inference.
While MCMC methods draw a sample from the joint posterior distribution,
the Laplace approximation is a method that approximates posterior distributions of the model parameters to Gaussians, which is computationally more effective.

Within the INLA framework, the posterior distribution of the latent 
Gaussian variables $\pmb{x}$ and hyperparameters of the model $\pmb{\theta}$ is:

\begin{align}
   \pi(\pmb{x}, \pmb{\theta} ~|~ \pmb{y}) = \frac{
    \pi(\pmb{y} ~|~  \pmb{x},  \pmb{\theta})~ 
   \pi(\pmb{x}, \pmb{\theta})}{\pi(\pmb{y})} \propto 
   \pi(\pmb{y} ~|~  \pmb{x},  \pmb{\theta})~ 
   \pi(\pmb{x}, \pmb{\theta}),
    \label{joint_post}
\end{align}

\noindent where $\pmb{y}=(y_{1},...,y_{n})$ represents set of observations i.e. a pixel matrix.  
Every pixel is treated with a latent Gaussian effect, 
so that each 
$x_{i}$ (which is just mean value $\mu_{i}$ and
standard deviation $\sigma_{i}$ of a Gaussian for a pixel $y_{i}$) corresponds to observation $y_{i}$,
where $i\in[1,...,n]$. The observations are conditionally independent given the latent effect $\pmb{x}$ and the hyperparameters $\pmb{\theta}$, so the model likelihood is:

\begin{align}
 \pi(\pmb{y} ~|~  \pmb{x},  \pmb{\theta}) =
 \prod_{i} \pi(y_{i}  ~|~ x_{i}, \pmb{\theta}).
\end{align}

\noindent The joint distribution of the latent effects and the hyperparameters, $\pi(\pmb{x}, \pmb{\theta})$ can be written as 
$\pi(\pmb{x}  ~|~ \pmb{\theta})~\pi(\pmb{\theta})$, where 
$\pi(\pmb{\theta})$ represents the prior distribution of hyperparameters $\pmb{\theta}$. 
It is assumed that the spatial information can be treated as a discrete sampling of an underlying continuous spatial field, a latent Gaussian Markov random field (GMRF), that takes into account the spatial correlations, and whose hyperparameters are inferred in the process.
For a GMRF, the posterior distribution of the latent effects is:
\begin{align}
 \pi(\pmb{x} ~|~  \pmb{\theta}) \propto
 |\pmb{Q}(\pmb{\theta})|^{1/2} \exp { \textlbrackdbl  -\frac{1}{2}
 \pmb{x}^{T}\pmb{Q}(\pmb{\theta})~\pmb{x}
 \textrbrackdbl},
\end{align}

\noindent where $\pmb{Q}(\pmb{\theta})$ represents a precision matrix, or inverse of a covariance matrix, which depends on a vector of hyperparameters $\pmb{\theta}$. This kernel matrix is what actually treats the spatial correlation between neighbouring pixels.
Using equation \ref{joint_post}, the joint posterior distribution of the latent effects and hyperparameters can be written as:

\begin{align}
  \begin{aligned}
  \pi(\pmb{x}, \pmb{\theta} ~|~ \pmb{y})\propto
  \pi(\pmb{\theta}) |\pmb{Q}(\pmb{\theta})|^{1/2} \exp { \textlbrackdbl  -\frac{1}{2}
 \pmb{x}^{T}\pmb{Q}(\pmb{\theta})~\pmb{x}
 \textrbrackdbl}  \prod_{i} \pi(y_{i}  ~|~ x_{i}, \pmb{\theta}) \\ 
 =\pi(\pmb{\theta}) |\pmb{Q}(\pmb{\theta})|^{1/2} \exp { \textlbrackdbl  -\frac{1}{2}
 \pmb{x}^{T}\pmb{Q}(\pmb{\theta})~\pmb{x} +
 \sum_{i} \log (\pi(y_{i}  ~|~ x_{i}, \pmb{\theta}))
 \textrbrackdbl}.
 \label{post_joint}
  \end{aligned}
\end{align}

\noindent Instead of obtaining the exact posterior distribution from
equation \ref{post_joint}, INLA
approximates the posterior marginals of the latent effects and hyperparameters. Thus, the key feature of INLA methodology is 
to use appropriate approximations for the following integrals:

\begin{align}
 \pi( x_{i} ~|~ \pmb{y})=\int \pi( x_{i} ~|~ \pmb{\theta}, \pmb{y})
 \pi( \pmb{\theta}~|~ \pmb{y}) ~ d\pmb{\theta}
\end{align}

\begin{align}
 \pi( \theta_{j} ~|~ \pmb{y})=\int
 \pi( \pmb{\theta}~|~ \pmb{y}) ~ d\pmb{\theta}_{-j}, 
\end{align}

\noindent where $\pmb{\theta}_{-j}$ is a vector of hyperparameters
$\pmb{\theta}$  without element $\theta_{j}$. INLA constructs nested approximations:

\begin{align}
 \tilde{\pi}( x_{i} ~|~ \pmb{y})=\int \tilde{\pi}( x_{i} ~|~ \pmb{\theta}, \pmb{y})
 \tilde{\pi}( \pmb{\theta}~|~ \pmb{y}) ~ d\pmb{\theta}
   \label{post_lat}
\end{align}

\begin{align}
 \tilde{\pi}( \theta_{j} ~|~ \pmb{y})=\int
 \tilde{\pi}( \pmb{\theta}~|~ \pmb{y}) ~ d\pmb{\theta}_{-j},
\end{align}

\noindent where $\tilde{\pi}(\cdot~|~\cdot)$  is an approximated posterior density. 

Using the Laplace approximation, the posterior marginals of hyperparameters
$\pi( \pmb{\theta}~|~ \pmb{y})$ at a specific value  $\pmb{\theta}=\pmb{\theta}_{j}$ can be written as:

\begin{align}
 \tilde{\pi}( \pmb{\theta}_{j}~|~ \pmb{y})\propto \frac{\pi(\pmb{x},\pmb{\theta}_{j},\pmb{y})}
 {\tilde{\pi}_{G}(\pmb{x}~|~\pmb{\theta}_{j},\pmb{y})}  \propto 
 \frac{\pi(\pmb{y}~|~\pmb{x},\pmb{\theta}_{j})
 \pi(\pmb{x}~|~\pmb{\theta}_{j}) \pi(\pmb{\theta}_{j})
 }
 {\tilde{\pi}_{G}(\pmb{x}~|~\pmb{\theta}_{j},\pmb{y})} \mid_{\pmb{x}=\pmb{x}^{*}(\pmb{\theta}_{j})}  
 ,  
 \label{post_hyper}
\end{align}

\begin{align}
 \tilde{\pi}_{G}(\pmb{x}~|~\pmb{\theta},\pmb{y}) \propto  
 \exp { \textlbrackdbl  -\frac{1}{2}
 \pmb{x}^{T}\pmb{Q}(\pmb{\theta})~\pmb{x} +
 \sum_{i} g_{i}(x_{i})
 \textrbrackdbl}
\end{align}

\noindent where $\tilde{\pi}_{G}(\pmb{x}~|~\pmb{\theta},\pmb{y})$
is the Gaussian approximation to the full conditional of $\pmb{x}$ , and $\pmb{x^{*}}(\pmb{\theta}_{j})$ is the mode
of the full conditional $\pmb{x}$  for given $\pmb{\theta}_{j}$ (\citealt{inla}). 

The posterior marginals of the latent effects are numerically integrated as following:

\begin{align}
\tilde{\pi}( x_{i} ~|~ \pmb{y}) \backsimeq 
\sum_{j} \tilde{\pi}( x_{i} ~|~ \pmb{\theta}_{j}, \pmb{y})
 \tilde{\pi}( \pmb{\theta}_{j}~|~ \pmb{y}) \Delta_{j},
\end{align}

\noindent where $\Delta_{j}$ represents integration weight. 
However, a good approximation for $\tilde{\pi}( x_{i} ~|~ \pmb{\theta}, \pmb{y})$
is required and INLA offers three different options:
Gaussian approximation, Laplace approximation and simplified Laplace approximation (\citealt{inla}). In this work we used
the simplified Laplace approximation, which is the default one 
and represents a compromise between the accuracy of the Laplace approximation and the reduced computational cost achieved with the Gaussian approximation.

As described above INLA uses spatial correlations between data points to reconstruct missing and/or noisy data.
We refer to  \cite{inla} and \citet{Gomez-Rubio}
for more details on the mathematical background of INLA  and the methods it employs. 

INLA is available as R-INLA package\footnote{\url{http://www.r-inla.org/}}
designed for modeling spatial data (\citealt{r-inla}). 
In conjunction with INLA, we use the PARDISO package (\citealt{PARDISO}) which represents a high-performance software for solving sparse symmetry that arises in the R-INLA approach to Bayesian inference, and thus reduces INLA's computational time.
Since synthetic observations simulated with the \textsc{skirt} code represent 
non-random spatial structures, we explore the potential of INLA method as a tool for enhancing the MCRT images, and thus, optimizing the total computational run time invested in obtaining high quality images.

\subsection{Dimensionality reduction}

In order to achieve additional reduction in computational time,
we implement dimensionality reduction techniques in our methodology for \textsc{skirt} code post-processing.
Dimensionality reduction methods diminish the number of attributes in an original dataset while attempting to preserve as much of the variation  as possible.

\subsubsection{PCA}

Principal component analysis (PCA) is an unsupervised method that analyses the feature space of the training data and creates orthogonal vectors (linear combinations of the initial variables) whose direction indicates the most variability (for derivations see \citealt{PCA1}; \citealt{PCA2}; for a modern review see e.g. \citealt{PCA3}). These new vectors in the transformed data set 
are called eigenvectors, or principal components, while eigenvalues
represent the coefficients attached to eigenvectors, which give the relative amount of variance carried in each principal component.
Principal components are uncorrelated and most of the information is compressed into the first components. Even though the transformed dataset has the same number of dimensions as the original dataset, by discarding components with low information, dimensionality reduction is achieved.

Presently, within the astronomy and cosmology fields of research, PCA is mostly used in algorithms and/or pipelines that implement different combinations of learning methods (\citealt{UMT}; \citealt{HC5}).
In this work we use \texttt{prcomp}\footnote{\url{https://www.rdocumentation.org/packages/stats/versions/3.6.2/topics/prcomp}} R function.

\subsubsection{NMF}

Non-negative matrix factorization (NMF) is another unsupervised dimensionality reduction method widely used in various fields
(e.g. \citealt{Lee}; \citealt{Tandon}). Similar to PCA, NMF decomposes the feature space of the training dataset into the product of two matrices with smaller ranks. One matrix represents the NMF features that are linear combinations of the original dataset features, while the second NMF matrix contains coefficients i.e. weights associated with NMF features. NMF is initiated by attributing guess values to the elements of the reduced rank matrices which are then iteratively updated until both NMF matrices are stable and their product approaches the original training dataset. Although very similar to PCA, NMF has an additional requirement that both original and decomposed matrices have no negative elements. This property might be beneficial when treating data  with only non-negative values, such as astronomical fluxes (\citealt{nmf1}; \citealt{nmf2}). Here we employ \texttt{nmf}\footnote{\url{https://www.rdocumentation.org/packages/NMF/versions/0.24.0/topics/nmf}} R function.

\subsection{Implementation}

\begin{figure*}
   \resizebox{\hsize}{!}
            {\includegraphics[width=.3\textwidth]{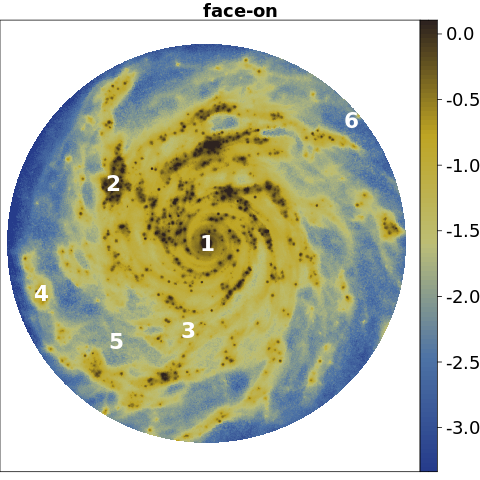}\hfill
\includegraphics[width=.3\textwidth]{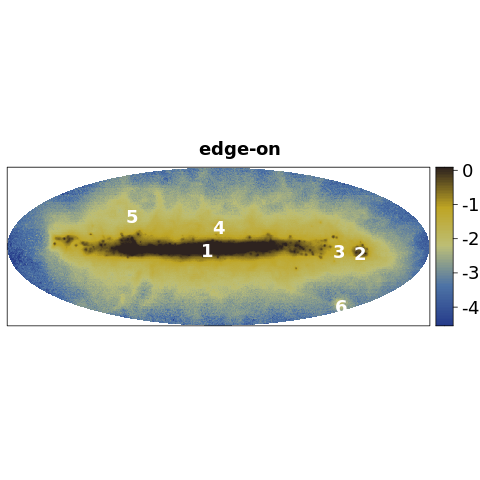}\hfill
\includegraphics[width=.3\textwidth]{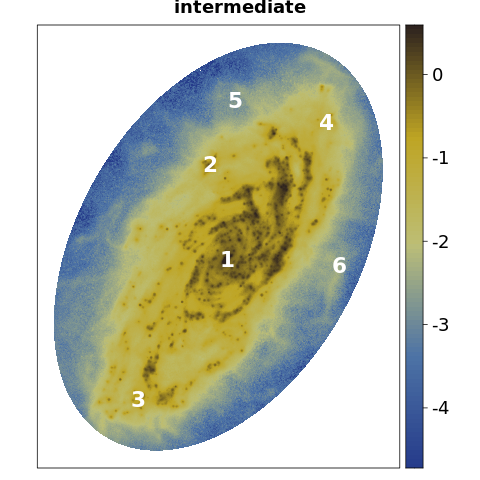}}
      \caption{HPN reference images of Au-16 galaxy
      at $\lambda=7.88~\mu$m, viewed at 'face-on' (right), 'edge-on' (middle) and 'intermediate' (right) angle. 
      Colour indicates flux density in MJy/sr , given in logarithmic scale. Numbers on spatial maps indicate positions of pixels whose SEDs are shown in section \ref{results_sed}. 
                    }
         \label{au16}
   \end{figure*}

We apply our method on Au-16 galaxy, which represents
a typical MW type galaxy with well resolved spiral structure.
Figure \ref{au16} shows Au-16 galaxy viewed from different angles, 'face-on', 'edge-on' and 'intermediate', at wavelength $\lambda=7.88~\mu$m.
These high resolution synthetic images are the result of a \textsc{skirt} simulation performed with $3\times10^{10}$ photon packages, hereafter referred to as the high photon number (HPN) reference images. 
The optimal number of photon packages used for HPN reference images
has been previously tested by \cite{Kapoor}, based on pixel-to-pixel relative error 
calculation. The authors have shown that 
simulations performed with $3\times10^{10}$ photon packages
provide sufficiently high SNR for most of the bands, with the exception 
of certain low flux regions in UV and submm ranges.
Those images are available as part of the \textsc{skirt} Auriga project\footnote{\url{https://www.auriga.ugent.be/DataZenodoRedirect.html}}
and we will use them to evaluate the performance of our method.
In addition, we run \textsc{skirt} simulations of the same galaxy, but
with lower number of photon packages, 
$3\times10^{8}$ and $3\times10^{9}$, and use them as low photon number (LPN) input images.
LPN input images require only $\sim2\%$ and $\sim11\%$  of the HPN reference simulation execution time, for 
$3\times10^{8}$ and $3\times10^{9}$ photon packages, respectively.
This time scaling justifies the implementation of the here proposed methodology, capable to reproduce high resolution synthetic observations using  time effective LPN images as an input.

Even though INLA provides a promising path to overcome the computational limits of \textsc{skirt} simulations,
INLA reconstructions alone can also be time costly, 
especially for large data cubes such as in this particular case.
The \textsc{skirt} simulation output of our dataset is a cube of 3072$\times$3072 pixels
at 50 wavelength bins, from $0.15~\mu$m to 1.24 mm.
Initial conditions for \textsc{skirt} Auriga simulations are taken from the Auriga simulation snapshots, using a cubical aperture to extract the data.
The aperture is centred at the galaxy centre, with a side length 
adjusted in a manner that it includes
most of the bound galaxy particles, but at the same time avoids
secondary structures, such as satellite galaxies. 
\cite{Kapoor} used the side length 
twice the radius at which the 'face-on' stellar surface density within 10 kpc of the mid plane falls bellow the threshold value 
$2\times10^{5}\Msun/\rm{kpc}^{2}$. 
Since the same cubic aperture is used for all three inclinations,
'edge-on' images will have extended regions with essentially no information. 
In post-processing we decided to exclude those outermost pixels, resulting with different cube dimensions:
2150$\times$2150$\times$50 for 'face-on'; 
2400$\times$900$\times$50 for 'edge-on' and 1800$\times$2200$\times$50 for 'intermediate' cube.

Full data cube reconstruction requires 50 individual 
spatial reconstructions, one for each of the wavelength bins.
Such reconstructions are labelled 'pure INLA' in the following text.
We employ PCA and NMF techniques to reduce the number of dimensions, i.e. number of individual spatial maps
to be reconstructed with INLA.
Both methods transform the original dataset in the spectral dimension, such that these spectra are described with a new orthogonal basis (in order of decreasing variability) instead of the 50 wavelengths. In this transformed space,
the first components will carry most of the information, while the rest
are responsible for less prominent features and noise,
and can be discarded in order to trade some of the precision
for efficiency. 
After performing PCA/NMF transformation on the initial variables (wavelength bins), these are replaced by a reduced number of principal components, whose maps are used as input to INLA. INLA then reconstructs the spatial maps of the coefficients
attached to our selected set of  principal components, instead of flux densities from
the original data cube. Reconstructed spatial maps are then 
multiplied by the transpose of the principal components, 
resulting in the reconstructed data cube with the original dimensions, labelled 'PCA/NMF+INLA' reconstructions.
In Appendix \ref{color_maps_components} we show
spatial maps of PCA/NMF components before and after INLA reconstructions.
In Appendix \ref{pure_reconstructions} we compare PCA/NMF reconstructions before and after applying INLA. Pure PCA/NMF reconstructions show higher residuals compared to PCA/NMF+INLA reconstructions.

Additionally, the INLA spatial reconstruction time can be reduced by 
sampling the input map data, instead of using complete spatial information. INLA shows optimal performance
when applied to sparse data, the execution time is notably reduced while the quality of reconstructions is influenced only to a lesser degree (\citealt{santiago}). This is further studied in section \ref{cut}, where we perform INLA reconstructions using different sampling percentages.
We also test different number of PCA/NMF components in order to find an optimal compromise between the quality of reconstructed maps and the required running times.

\section{Results and discussion}
\label{results}

\subsection{Field cuts optimization}
\label{cut}

We test the optimal pixel sampling percentage using the 600x600x50 cut of 'face-on' cube. 
The quality of INLA reconstructions, compared to HPN reference images, is quantified by the normalized residuals, calculated as:

\begin{align}
 Residuals~(\%) = |\frac{X^{\textquotesingle}-X}{X}|\times 100 \%
 \label{res_equatio}
\end{align}

\noindent where $X^{\textquotesingle}$ and  $X$ 
refer to INLA reconstruction and HPN reference images, respectively.

\begin{figure}
   \centering
   \includegraphics[width=\hsize]{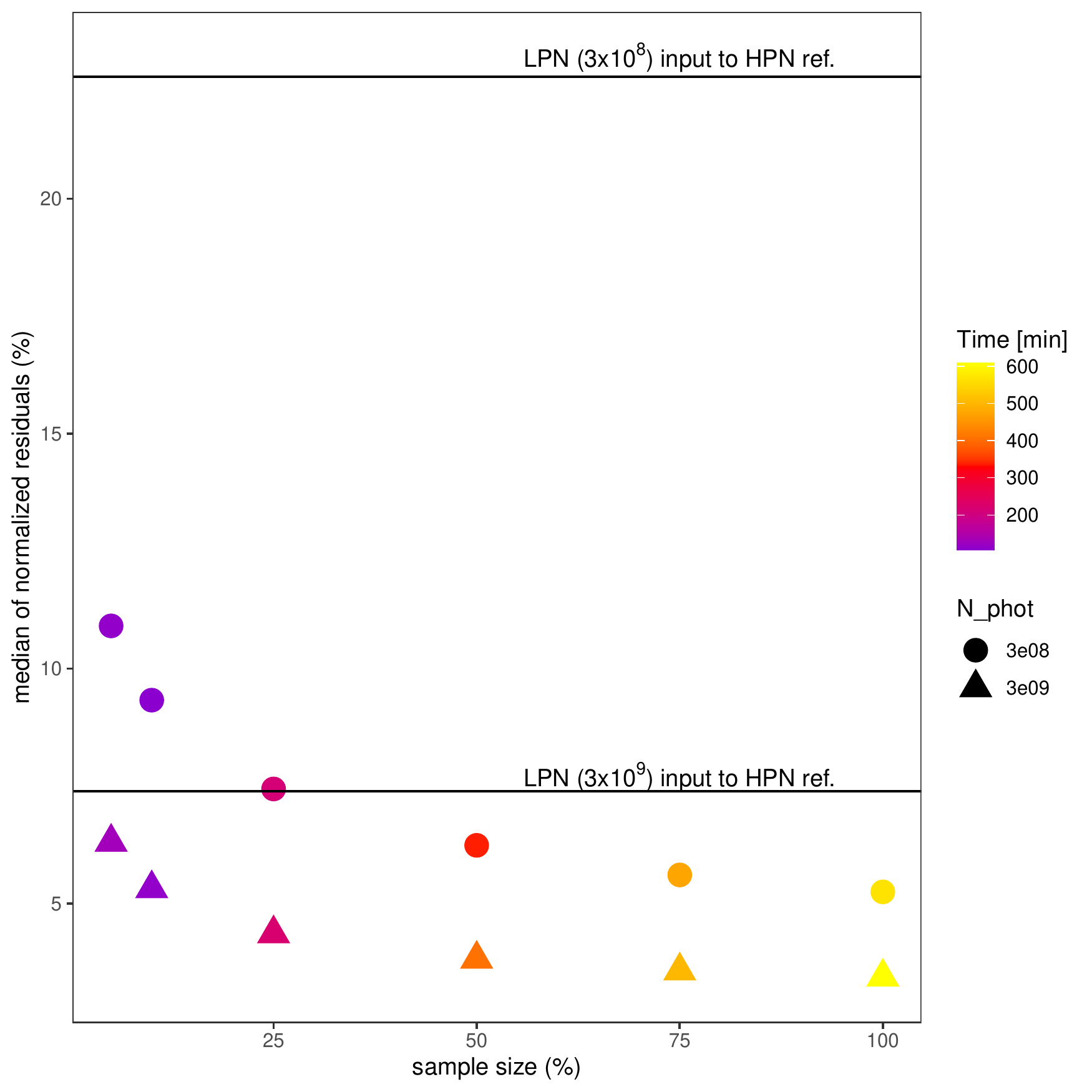}
      \caption{Median of normalized residuals of pure INLA reconstruction as a function of sampling percentage for 600x600x50 cut  of 'face-on' cube.
      Circles and triangles refer to $3\times10^{8}$
and $3\times10^{9}$ photon number realizations, respectively.
Horizontal lines refer to LPN input without spatial inference to HPN reference residuals.
The total running times are represented in different colours.}
         \label{pure_inla_res_cut}
   \end{figure}

Figure \ref{pure_inla_res_cut} presents the median of normalized residuals of pure INLA reconstructions as a function of the sampling
percentage used (5\%, 10\%, 25\%, 50\% and 100\%), for the 600x600x50 cut. The total running times for INLA reconstructions are represented with different colours. 
Horizontal lines refer to the median residuals of the LPN inputs in relation to the HPN reference image without applying INLA,
for $3\times10^{8}$ (circles) and $3\times10^{9}$ (triangles)
photon number simulations.

Increasing sample size improves the quality of reconstructions 
at the cost of a higher computational time.
In terms of balance between residuals and the running times,
INLA shows the optimal performances when applied to sparse data.
Figure \ref{pure_inla_res_cut} shows that using more than 25\% of the spatial data as INLA input significantly increases the running times with only a lesser ($1-2\%$) improvement in the quality of reconstructions. 
In the following sections, we present reconstructions using 
10\% and 25\% of the spatial information.

\begin{figure}
   \centering
   \includegraphics[width=\hsize]{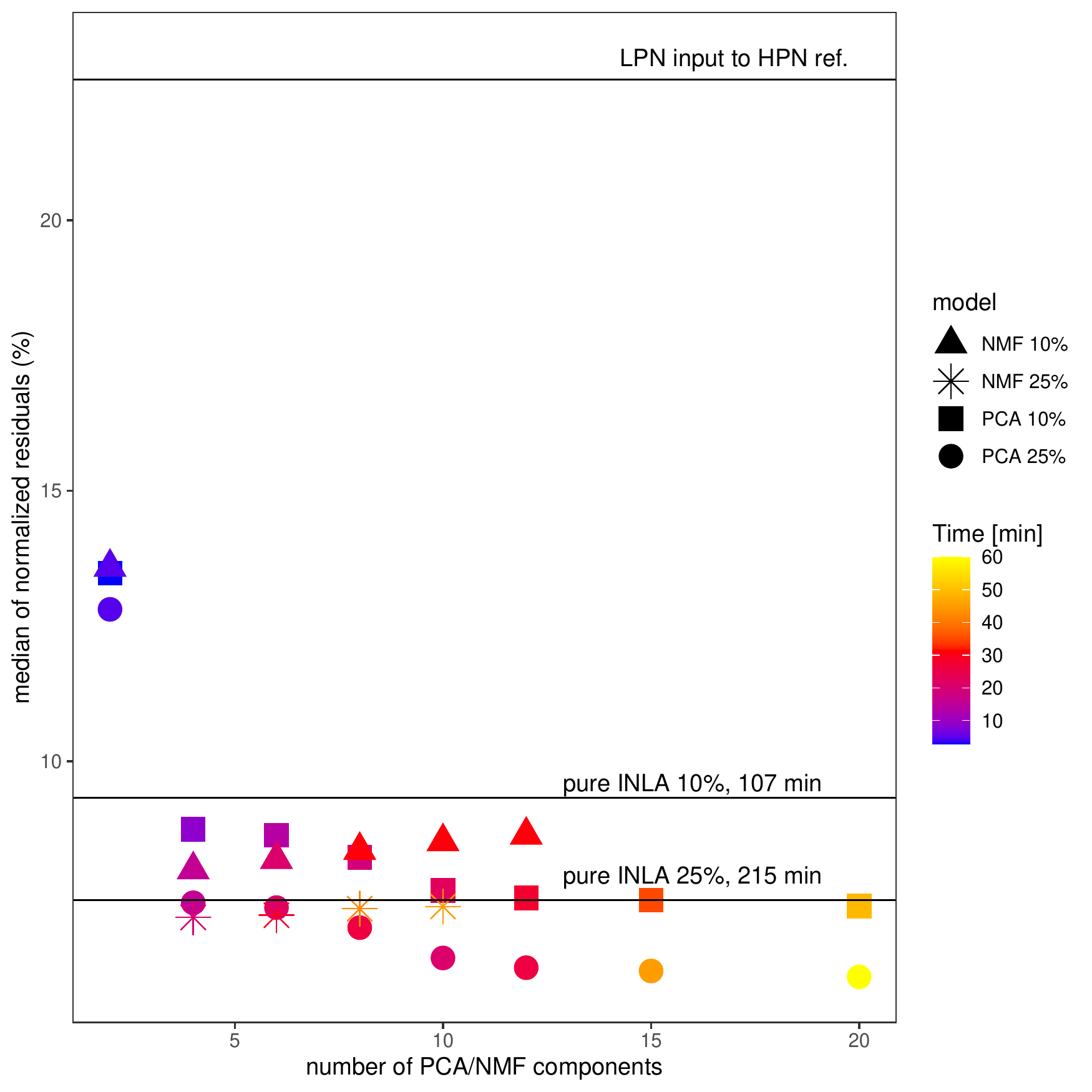}
      \caption{Median of normalized residuals of PCA/NMF+INLA reconstruction as a function of a number of PCA/NMF components used for the reconstruction, resulting from sampling 10\% and 25\% from $3\times10^{8}$ photon number realization.}
         \label{pca_res_cut}
   \end{figure}

Next, we explore the optimal number of PCA/NMF components.
PCA and NMF analyses are performed using 100\% of data within
600x600x50 cut, followed by sampling a percentage ( 10\% or 25\%) of spatial maps of PCA/NMF coefficients as an input to INLA.
Figure \ref{pca_res_cut} represents the median of normalized residuals
as a function of the number of PCA/NMF components used for the reconstruction.
The figure shows a clear improvement in reconstructions when more than two components are used, followed by a more gradual change. This behaviour suggests that the first components describe the overall structure and major features, while others represent e.g. diffuse emission and low flux regions.
In the case of PCA, a small decrease in residuals is noticed up to the tenth component while using more than ten PCA/NMF components does not improve the results. 
However, NMF does not show such a clear trend and using between 4 and 12 components results in only $\sim0.5\%$ change in median residuals.

Application of our methodology to this field cut shows that
using dimensionality reduction techniques prior to INLA 
drastically reduces the total computation time of the proposed post-processing technique without loss of precision. 
Additionally, even the quality of reconstructions is improved compared to pure INLA (horizontal lines in Figure \ref{pca_res_cut}), if more than two components are used. 

Next, we move to the full field processing.
Reconstructions of full cube images shown in the following section are performed using 10 PCA/NMF components, with data sampling of 10\% and 
25\%.

\subsection{Full field processing}
\label{results_full}

In this section, we present the results for full field
reconstructions using pure INLA and PCA/NMF+INLA techniques. 
We test our method using LPN \textsc{skirt} simulations ($3\times10^{8}$ and $3\times10^{9}$ photon packages) with different
inclination angles: 'face-on', 'edge-on' and 'intermediate'. 
INLA reconstructions are performed using 10\% or 25\% of the available spatial information. All figures shown in this section refer to reconstructions 
using 25\% of LPN ($3\times10^{8}$) input data, while the complete statistics of full field processing are summarised in Table \ref{results_table_int} and Table \ref{results_table}.

\subsubsection{Spectral reconstructions}
\label{results_sed}

\begin{figure*}
\begin{multicols}{3}\hspace{1em} 
    \includegraphics[width=1\linewidth]{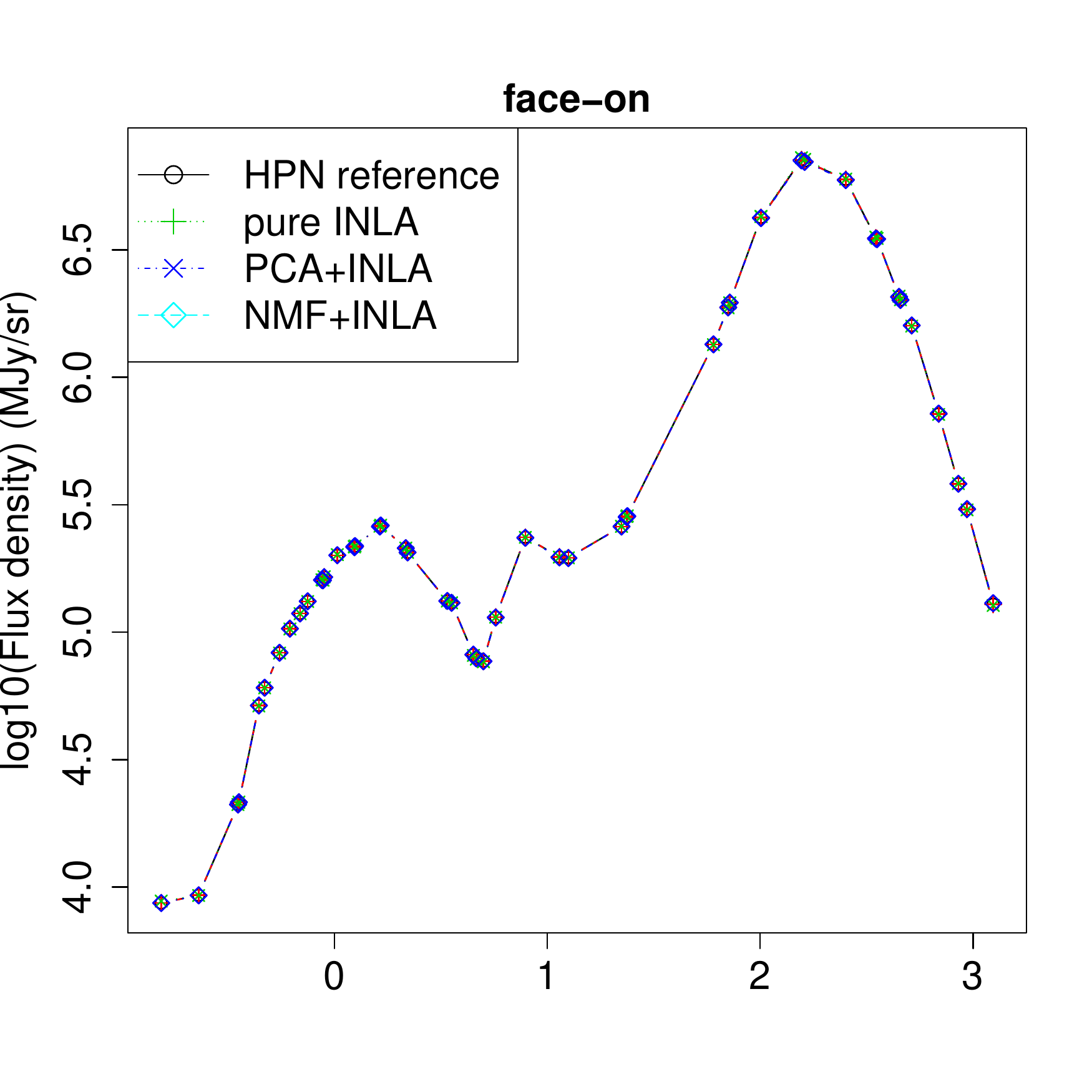}\par
    \includegraphics[width=1\linewidth]{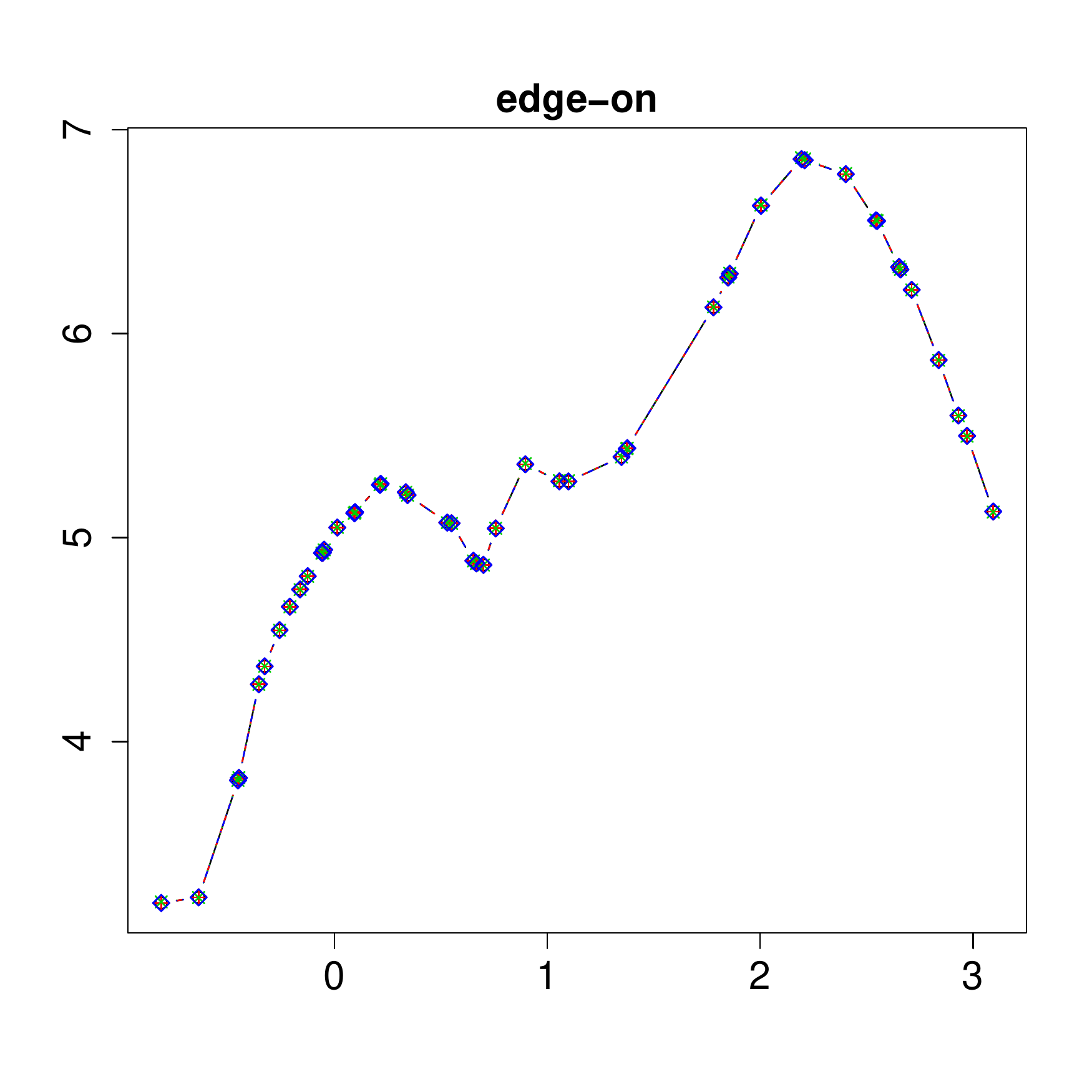}\par\hspace{-1em}
    \includegraphics[width=1\linewidth]{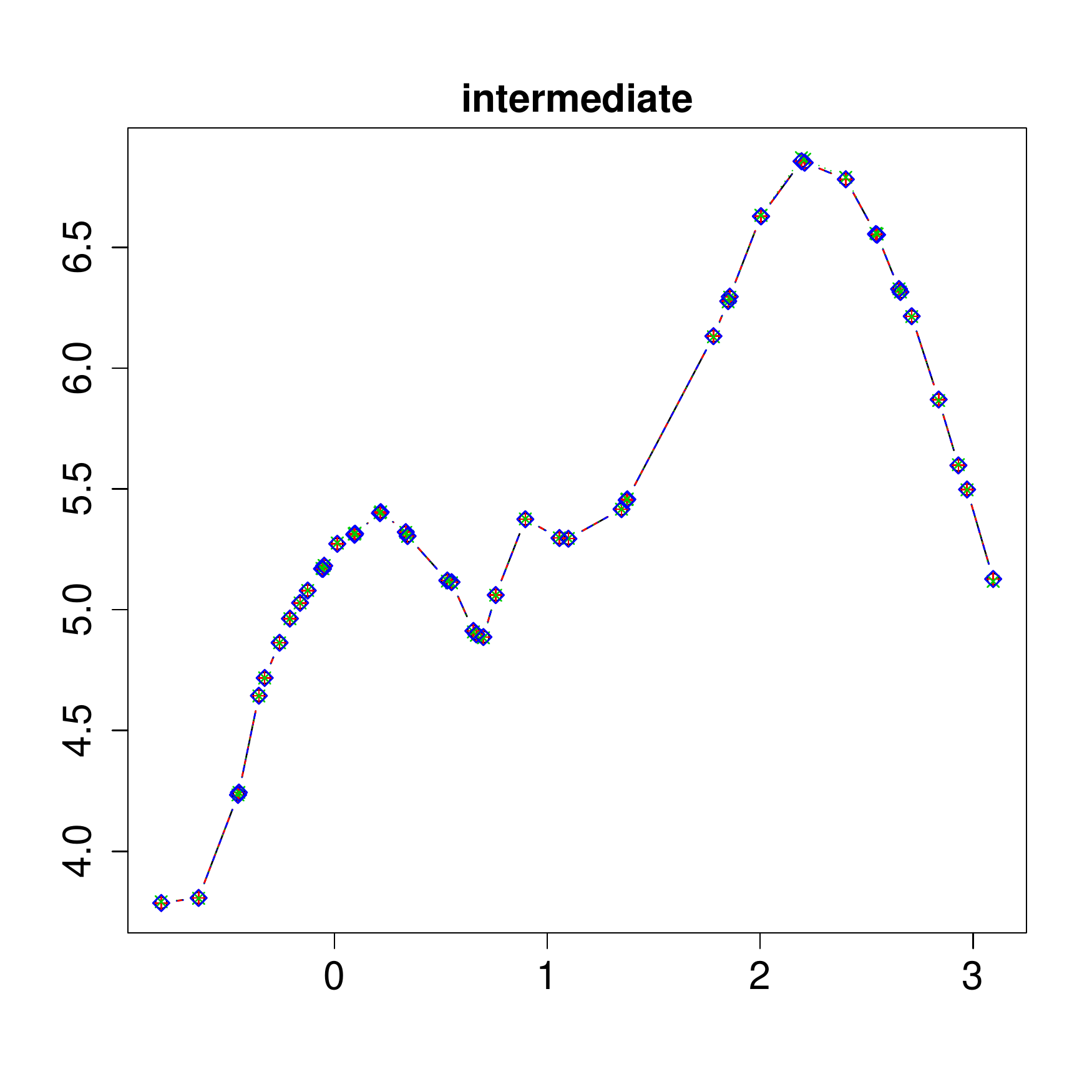}\par

    \end{multicols}
    \vspace{-5.5em}
\begin{multicols}{3}\hspace{1em}
    \includegraphics[width=1\linewidth]{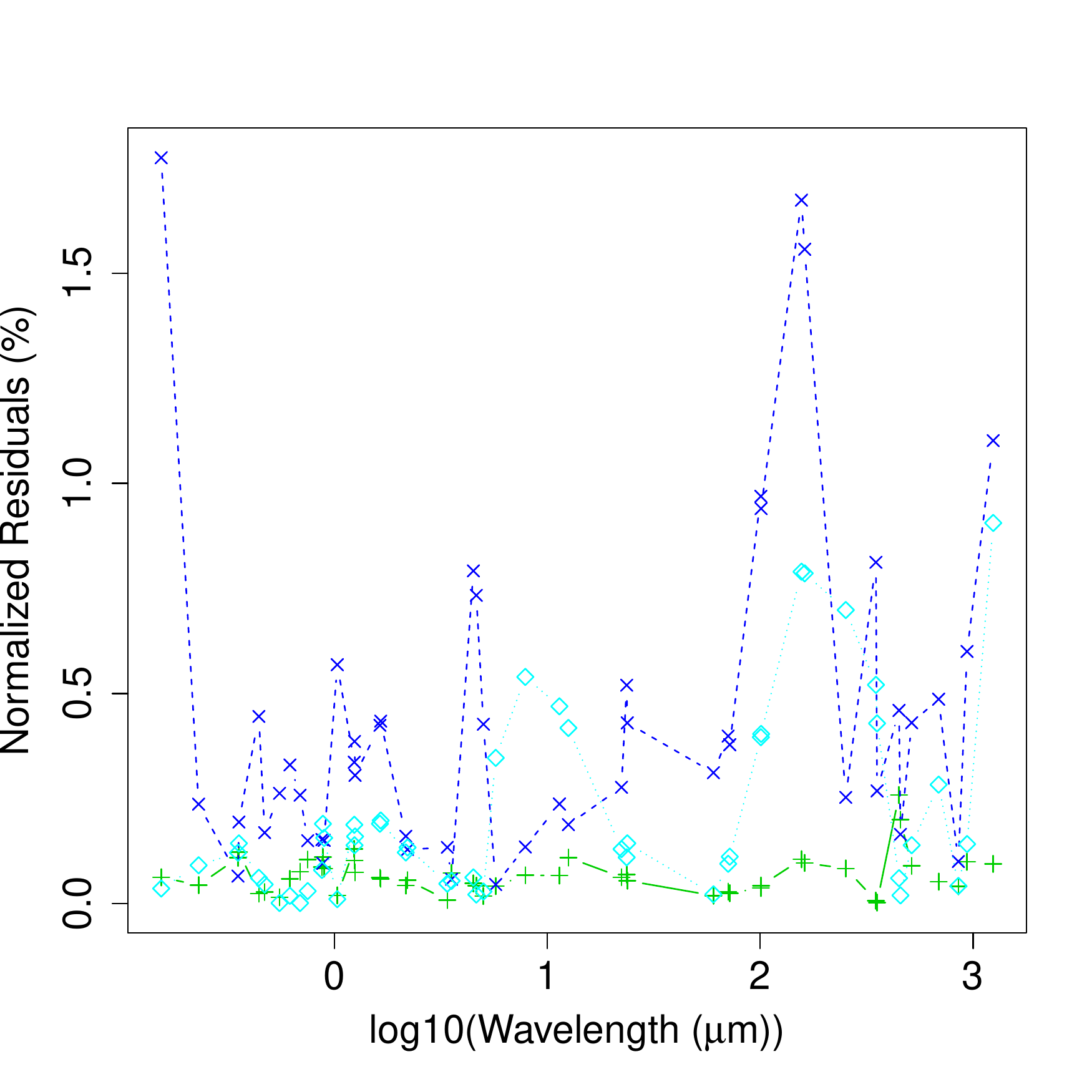}\par
    \includegraphics[width=1\linewidth]{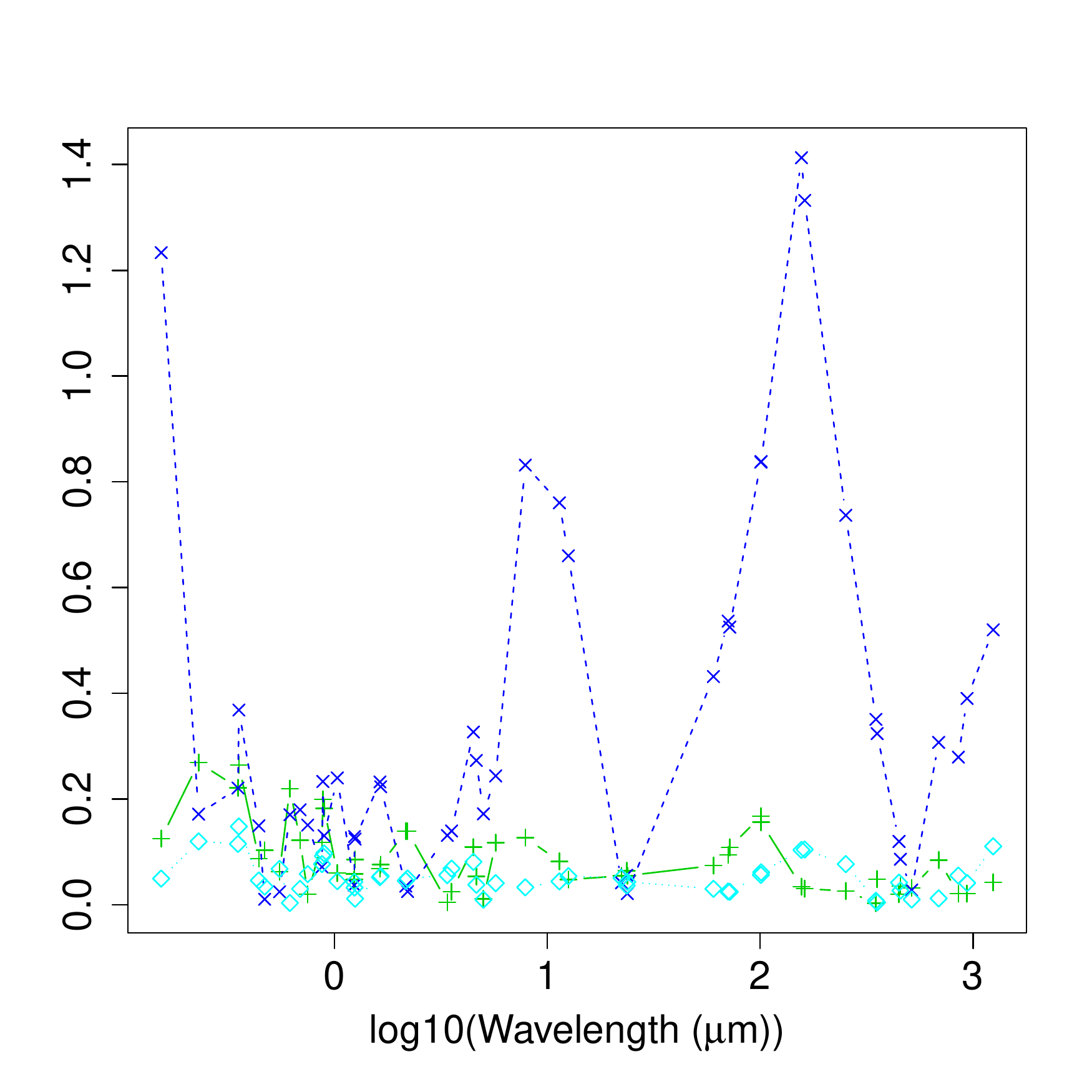}\par\hspace{-1em}
    \includegraphics[width=1\linewidth]{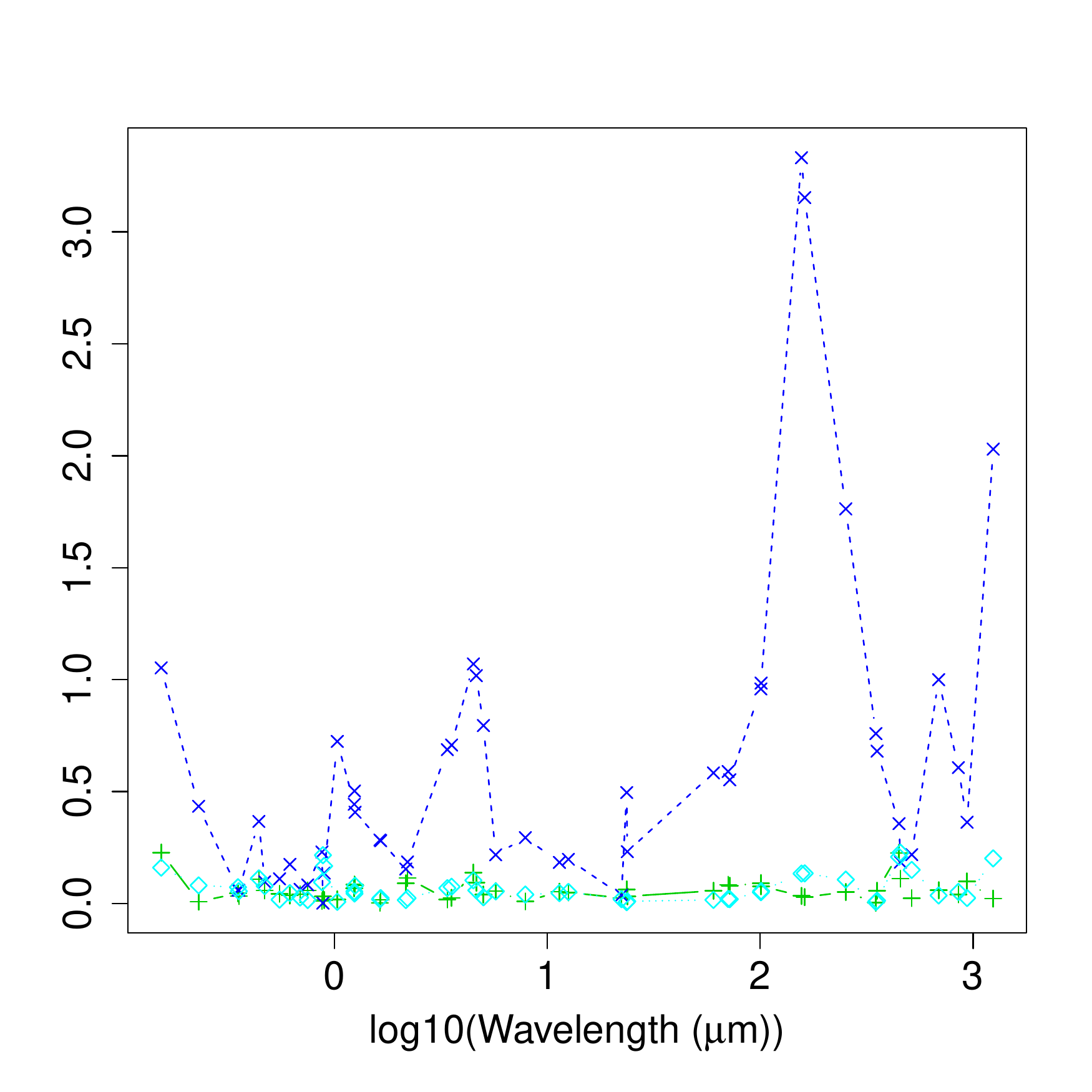}\par

\end{multicols}
      \caption{Integrated SEDs for HPN reference, pure INLA and  PCA/NMF+INLA reconstructions (upper panels) with the associated residuals
(lower panels),  for 'face-on' (left), 'edge-on' (middle) and
'intermediate' (right) cubes.} 
         \label{integrated_sed}
\end{figure*}

In this section, we explore how the reconstructed images follow  the expected SED of the HPN reference.

First, we compare the spatially integrated SEDs, obtained
from integrating all flux at each wavelength bin.
Figure \ref{integrated_sed} shows integrated SEDs for
HPN reference (\textopenbullet), pure INLA (+), PCA+INLA (\texttimes) and NMF+INLA ($\Diamond$) (upper panels) with the associated residuals
(lower panels),  for 'face-on' (left), 'edge-on' (middle) and
'intermediate' (right) cubes. For each of the employed techniques
integrated SEDs provide a good match to the HPN reference, 
throughout the whole wavelength range.
PCA+INLA reconstructions display the highest deviations
from the expected SEDs at wavelengths around $100~\mu$m, 
up to $\sim$1.5 \% for 'face-on' and 'edge-on' cubes, i.e.  
$\sim$ 3\% for the 'intermediate' cube.

Table \ref{results_table_int} summarises the statistics for 
all of the employed realizations: 'face-on', 'edge-on' and 'intermediate' cubes, simulated using  $3\times10^{8}$ and $3\times10^{9}$ photon packages. Median of the normalized residuals percentages are shown for 
pure INLA and PCA/NMF+INLA integrated SEDs, compared to HPN reference.
Predicted integrated SEDs are in excellent agreement with the expectations, with the median of the normalized residuals 
$\lesssim$ 0.1 \% for pure INLA and NMF+INLA reconstructions, and 
$\lesssim$ 0.3 \% for PCA+INLA.

\begin{table}[]
\caption{Median of the normalized residuals for integrated SEDs 
from pure INLA and PCA/NMF+INLA reconstructions, using different sampling percentages (10\% and 25\%) of $3\times10^{8}$ and $3\times10^{9}$ photon number cubes. }     
\centering
\label{results_table_int} 
\begin{tabular}{lcclll}
\multicolumn{1}{c}{}      & Photon               &                            & \multicolumn{3}{c}{Median of res.  (\%)}                                       \\
Angle                     & Number               & \multicolumn{1}{l}{sample} & \multicolumn{1}{c}{pure} & \multicolumn{1}{c}{PCA+} & \multicolumn{1}{c}{NMF+} \\
                          & Input                & \multicolumn{1}{l}{}       & \multicolumn{1}{c}{INLA} & \multicolumn{1}{c}{INLA} & \multicolumn{1}{c}{INLA} \\ \hline
\multicolumn{1}{c}{}      & $3\times10^{9}$      & 25\%                       & 0.02                     & 0.15                     & 0.03                     \\
\multicolumn{1}{c}{face-} &                      & 10\%                       & 0.04                     & 0.15                     & 0.07                     \\
\multicolumn{1}{c}{on}    & $3\times10^{8}$      & 25\%                       & 0.06                     & 0.32                     & 0.13                     \\
                          & \multicolumn{1}{l}{} & \multicolumn{1}{c}{10\%}   & 0.07                     & 0.30                     & 0.11                     \\ \cline{2-6} 
                          & $3\times10^{9}$      & 25\%                       & 0.03                     & 0.07                     & 0.03                     \\
edge-                     &                      & 10\%                       & 0.07                     & 0.13                     & 0.06                     \\
\multicolumn{1}{c}{on}    & $3\times10^{8}$      & 25\%                       & 0.07                     & 0.23                     & 0.05                     \\
                          & \multicolumn{1}{l}{} & \multicolumn{1}{c}{10\%}   & 0.17                     & 0.21                     & 0.06                     \\ \cline{2-6} 
                          & $3\times10^{9}$      & 25\%                       & 0.03                     & 0.22                     & 0.02                     \\
inter-                     &                      & 10\%                       & 0.04                     & 0.25                     & 0.03                     \\
mediate                        & $3\times10^{8}$      & 25\%                       & 0.05                     & 0.39                     & 0.05                     \\
                          & \multicolumn{1}{l}{} & \multicolumn{1}{c}{10\%}   & 0.08                     & 0.35                     & 0.08                     \\ \hline
\end{tabular}
\end{table}

\begin{figure*}
   \resizebox{\hsize}{!}{
\includegraphics[width=1\textwidth]{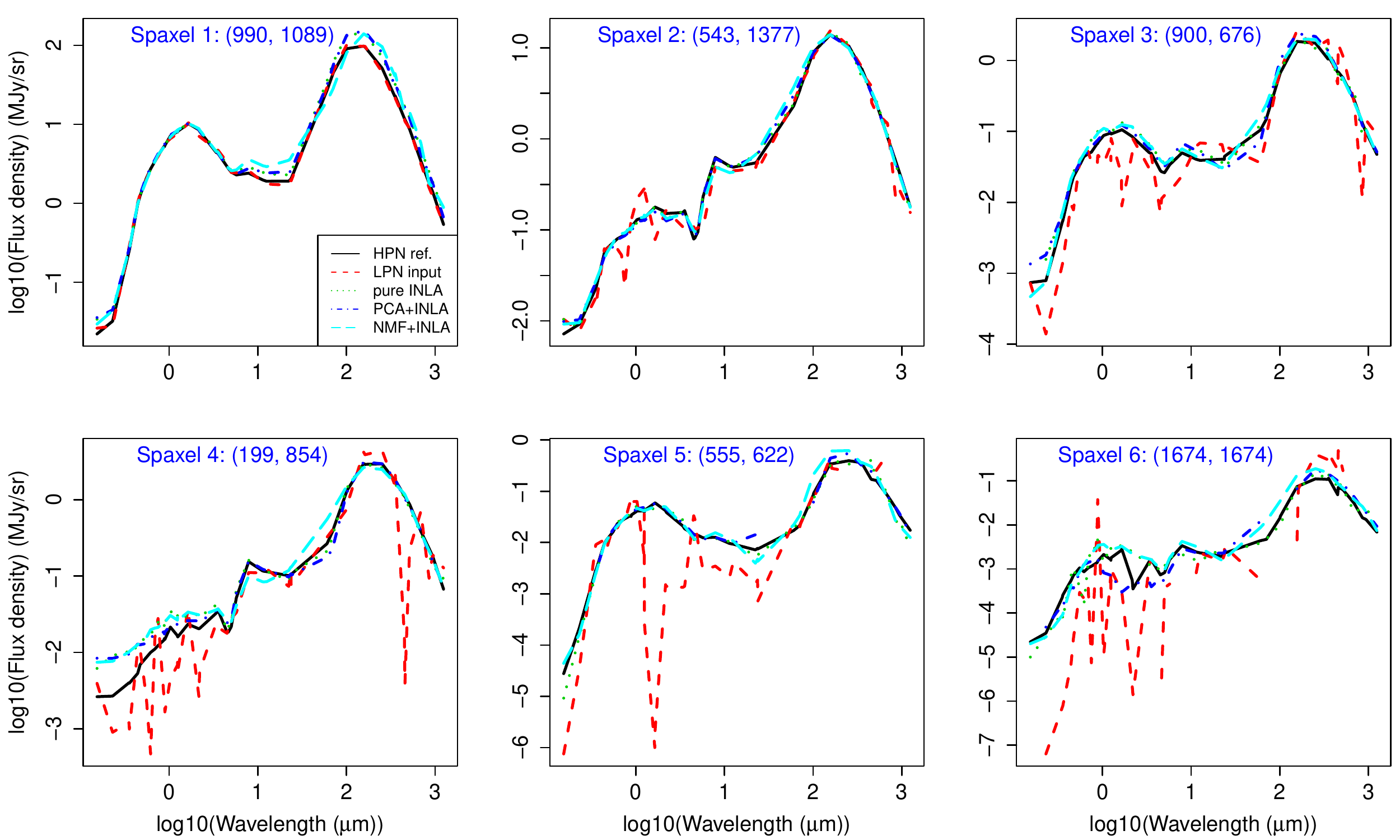}}
      \caption{Single spaxel SEDs for HPN reference (black) and LPN input
      (red) 'face-on' cubes, together with pure INLA (green) and PCA/NMF+INLA (blue and cyan) reconstructions. Spatial positions of the spaxels represented here are shown in Fig. \ref{au16}.}
         \label{spaxel_sed_face}
   \end{figure*}

\begin{figure*}
   \resizebox{\hsize}{!}{
\includegraphics[width=1\textwidth]{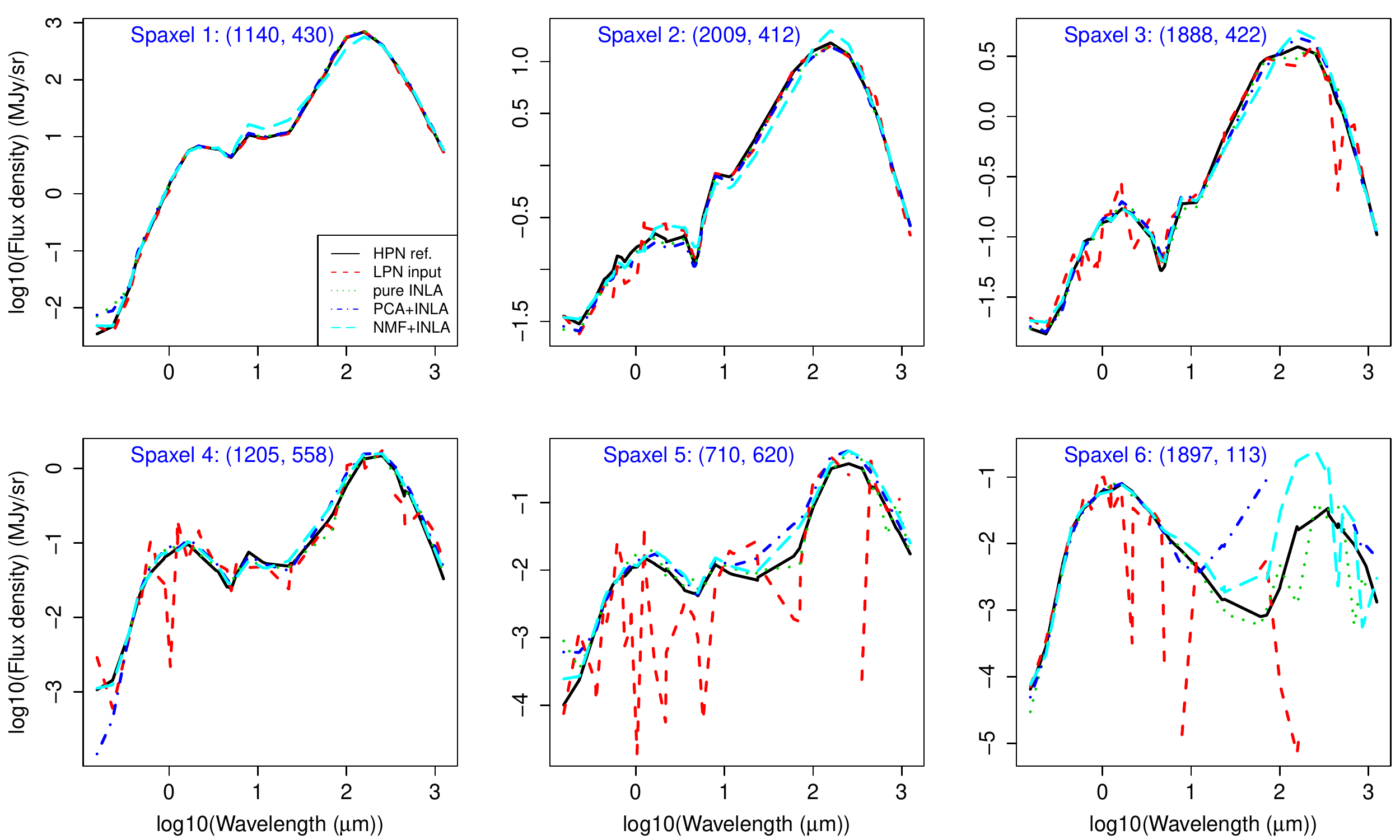}}
      \caption{Same as Fig. \ref{spaxel_sed_face} for 'edge-on' cube.}
         \label{spaxel_sed_edge}
   \end{figure*}

\begin{figure*}
   \resizebox{\hsize}{!}{
\includegraphics[width=1\textwidth]{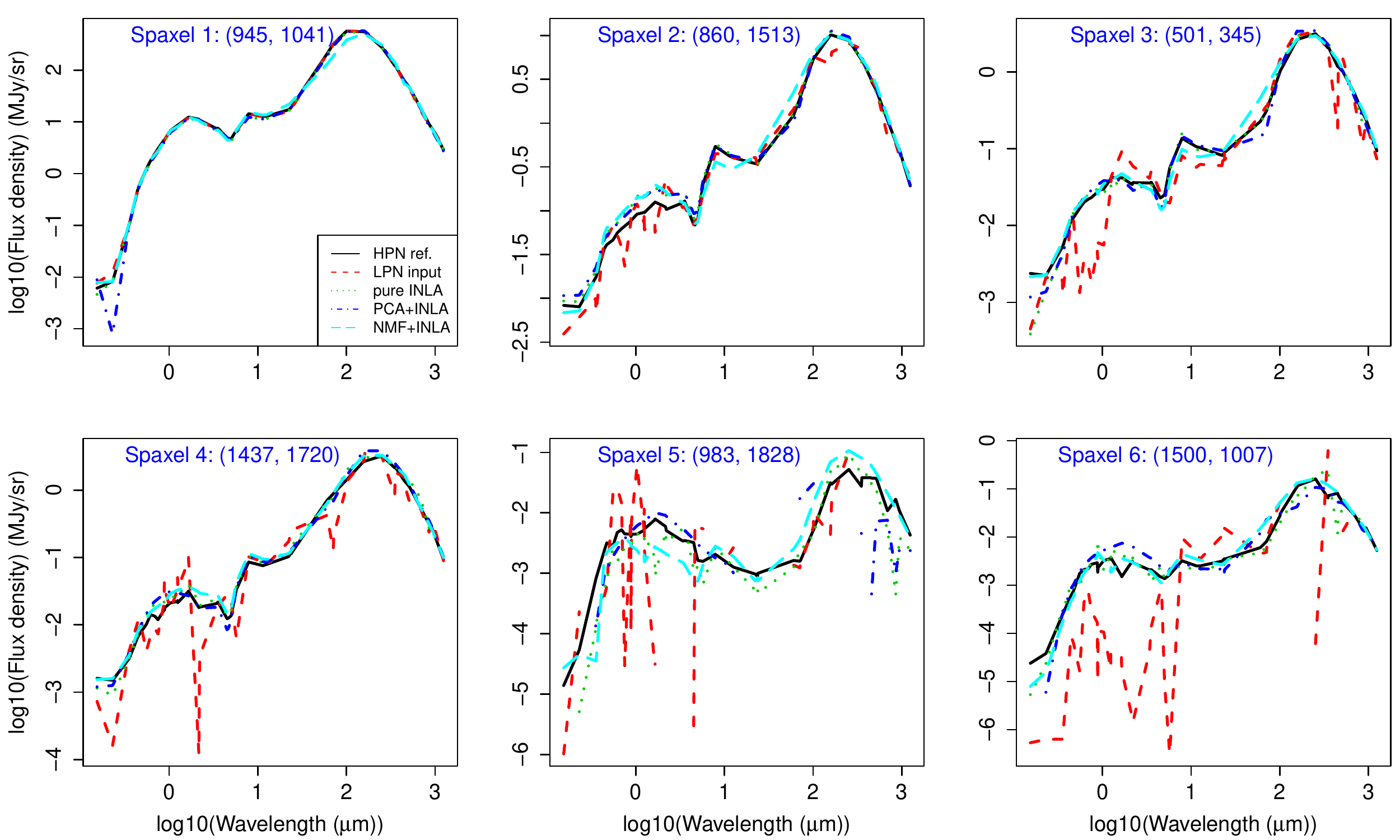}}
      \caption{Same as Fig. \ref{spaxel_sed_face} for 'intermediate' cube.}
         \label{spaxel_sed_random}
   \end{figure*}

Next, we explore how the quality of our predictions changes at various
regions throughout the galaxy plane. We inspect SEDs for individual spaxels\footnote{Spaxel refers to spectral pixel, with a spectrum 
associated to each spatial pixel.} 
whose spatial positions are marked in Figure \ref{au16}. 
The chosen spaxels occupy different regions of galaxy morphology, such as central parts, strong spiral arms, low flux density regions between spiral arms and galaxy outskirts.
Figures \ref{spaxel_sed_face}, \ref{spaxel_sed_edge} and \ref{spaxel_sed_random} show a single spaxel SEDs for 'face-on', 'edge-on' and 'intermediate' cubes, respectively, sampling 25\% of $3\times10^{8}$ photon number realization.

Both pure INLA (green) and PCA/NMF+INLA (blue and cyan lines)
reconstructions are in general in good agreement with 
HPN reference SEDs (black lines) for most pixel positions. 
The quality of reconstructions is correlated with 
the spaxel position and the flux density.
At high flux density regions, such as central parts and prominent spiral arms,
LPN input SEDs (red lines) closely follow the HPN reference,
and each of the employed techniques provides accurate predictions (spaxels 
at positions 1 and 2).
At lower flux densities along spiral arms (spaxels 3 and 4),
LPN input becomes noisy, however the quality of our reconstructions is not affected. 
However, reconstruction of faint, outermost parts of the galaxy (spaxels 5 and 6) becomes challenging given the LPN input SEDs with both noisy and missing data.
In these regions, pure INLA and NMF+INLA are overall able to recover the full SED of the HPN reference cube, while PCA+INLA occasionally fails in the reconstruction resulting in non-physical negative flux densities.
Such an example is shown in Figure \ref{spaxel_sed_edge},
where LPN input spaxel is not complete, meaning it has zero values at 
some wavelengths.
PCA+INLA reconstruction fails for spaxel 6,  positioned at the galaxy outskirt. NMF+INLA SED is complete, however the reconstructions overestimate the 
HPN reference SED, while pure INLA provides the best match.
In Table \ref{results_table} we provide summary of the median of the normalized residuals and the computation time of each methodology.

\subsubsection{Spatial reconstructions}

\begin{figure*}
   \resizebox{\hsize}{!}{
  \subfloat{\includegraphics{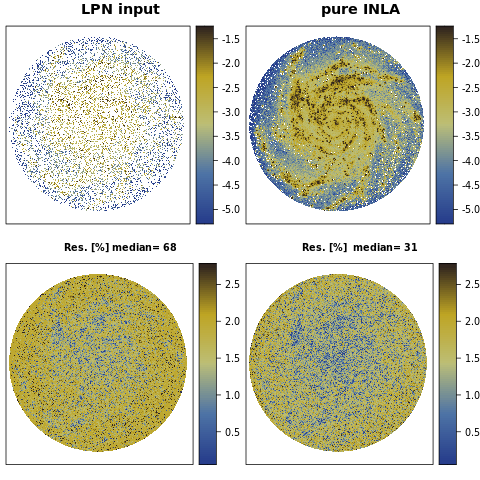} } \\
  \subfloat{\includegraphics{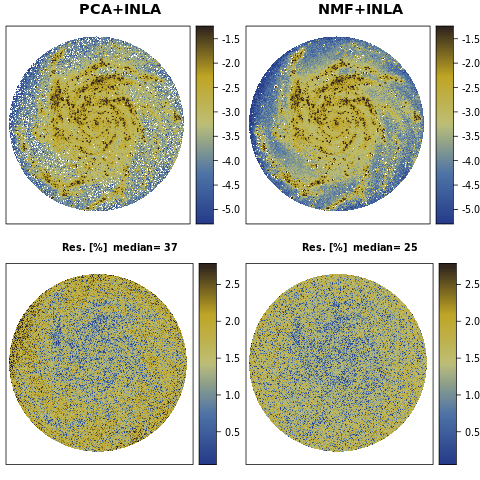} }

}

\caption{Spatial distribution of  LPN input cube and pure INLA, PCA/NMF+INLA reconstructions (upper panels)
with the associated residuals (bottom panels) at wavelength bin 0.23 $\mu$m. Sample size: 25\%, 'face-on' cube. Colour legends are given in logarithmic scale.
}
         \label{color_map2_face}
   \end{figure*}

\begin{figure*}
   \resizebox{\hsize}{!}{
  \subfloat{\includegraphics{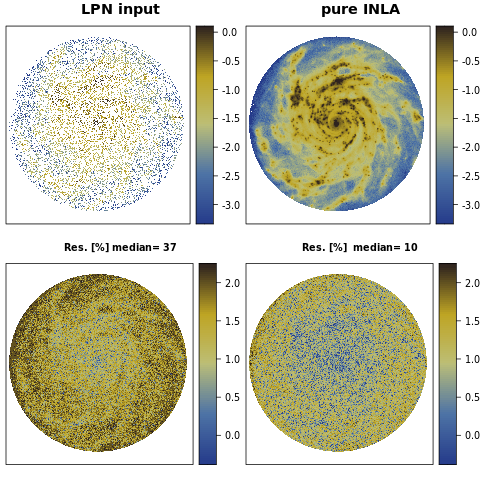} } \\
  \subfloat{\includegraphics{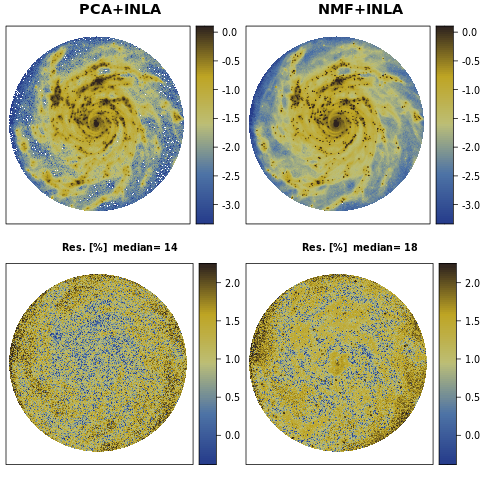} }

}

\caption{Same as Fig. \ref{color_map2_face} at wavelength bin 7.88 $\mu$m}
         \label{color_map28_face}
   \end{figure*}

\begin{figure*}
   \resizebox{\hsize}{!}{
  \subfloat{\includegraphics{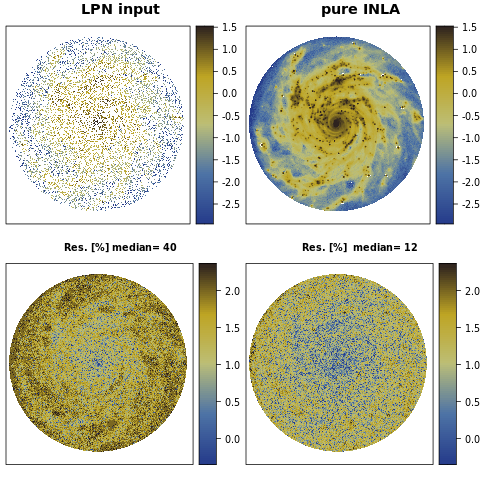} } \\
  \subfloat{\includegraphics{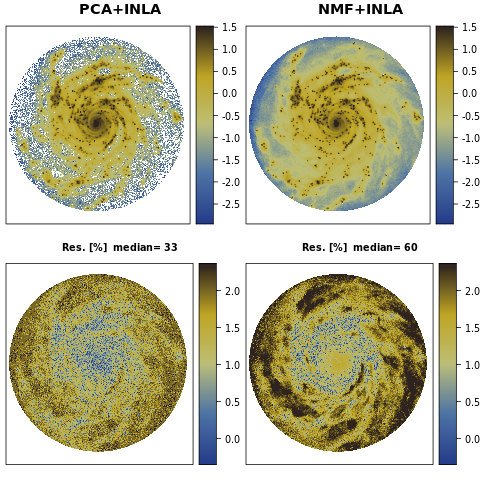} }

}

\caption{Same as Fig. \ref{color_map2_face} at wavelength bin 100.80 $\mu$m}
         \label{color_map37_face}
   \end{figure*}

\begin{figure*}
   \resizebox{\hsize}{!}{
  \subfloat{\includegraphics{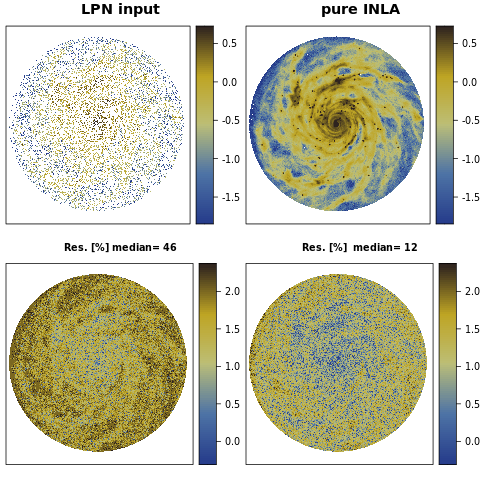} } \\
  \subfloat{\includegraphics{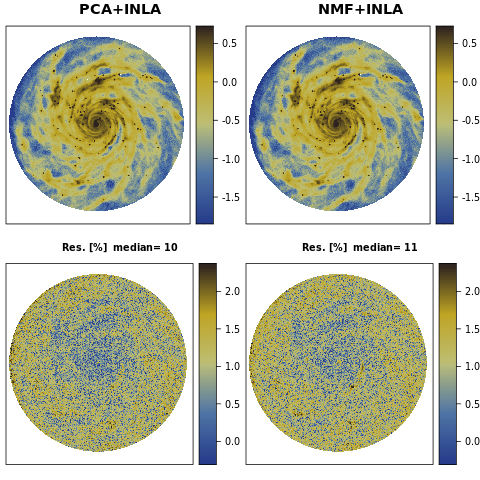} }

}

\caption{Same as Fig. \ref{color_map2_face} at wavelength bin 515.36 $\mu$m}
         \label{color_map46_face}
   \end{figure*}

\begin{figure*}
 \centering
  \includegraphics[width=.6\textwidth]{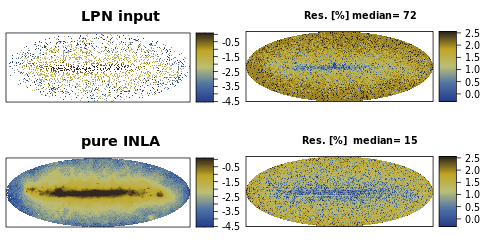} 
  \\
  \includegraphics[width=.6\textwidth]{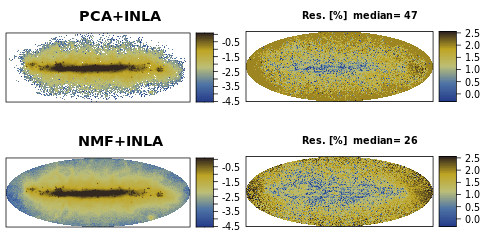}

\caption{Edge-on cube at wavelength bin 7.88 $\mu$m}
         \label{color_map28_edge}
   \end{figure*}

\begin{figure*}
   \resizebox{\hsize}{!}{
  \subfloat{\includegraphics{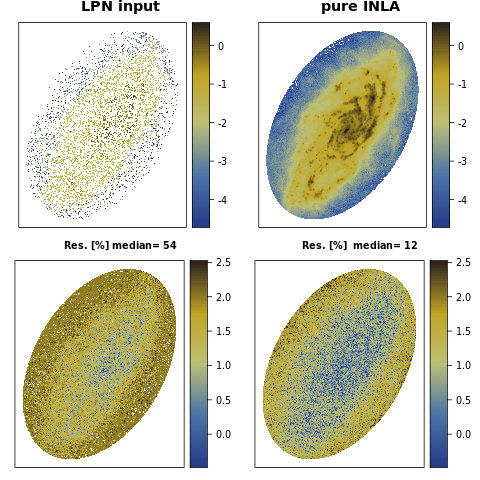} } \\
  \subfloat{\includegraphics{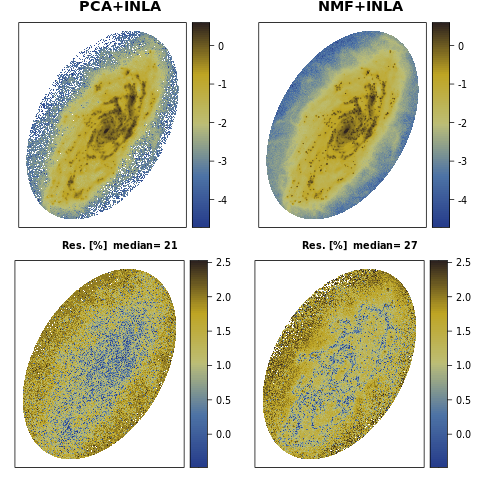} }

}

\caption{'Intermediate' cube at wavelength bin 7.88 $\mu$m}
         \label{color_map28_random}
   \end{figure*}

Figures in this section show spatial distribution of randomly sampling 25\% of
the LPN input cube spaxels, together with pure INLA and PCA/NMF+INLA reconstructions (upper panels). Each of these cubes is compared to the
HPN reference and normalized residuals are shown on the lower panels.
Figures \ref{color_map2_face}-\ref{color_map46_face} show
spatial reconstructions of the 'face-on' cube at four wavelength bins: 0.23 $\mu$m, 7.88 $\mu$m, 100.08 $\mu$m and $515.36~\mu$m. 
Figure \ref{color_map28_edge} and  Figure \ref{color_map28_random} show 'edge-on'  and 'intermediate cubes' at wavelength bin 7.88  $\mu$m.

By using a sample of LPN input cube our methods are able to recover underlying spatial information and reveal structures seen on HPN reference images (Figure \ref{au16}). 
The quality of reconstructions is
quantified by the median of normalized residuals (\%), calculated for each pixel at the given wavelength bin, and shown above each residual reconstruction image.
The pure INLA results display generally the lowest residuals, however this technique's extensive running time disqualifies it as a viable route to emulate MCRT codes. 
The PCA/NMF+INLA reconstructions have similar
residuals, with $\lesssim$ 25 \% of the running time required
for HPN reference (Table \ref{results_table}). However, at certain wavelength bins, the PCA+INLA technique fails in the reconstruction of $\sim10\%$
of spatial information, positioned at regions with the lowest flux density. 
Those regions with non-physical negative flux densities can be seen in
Figures \ref{color_map37_face}, \ref{color_map28_edge} and \ref{color_map28_random} as white pixels.
The problem with non-physical reconstructions induced by PCA analysis
can be avoided using NMF instead, since NMF 
enforces only positive elements in both original 
and decomposed matrices.
Using NMF+INLA method the problematic regions are reconstructed, but
with typically higher residuals compared to pure INLA.

\begin{figure*}
   \resizebox{\hsize}{!}{
\includegraphics[width=.5\textwidth]{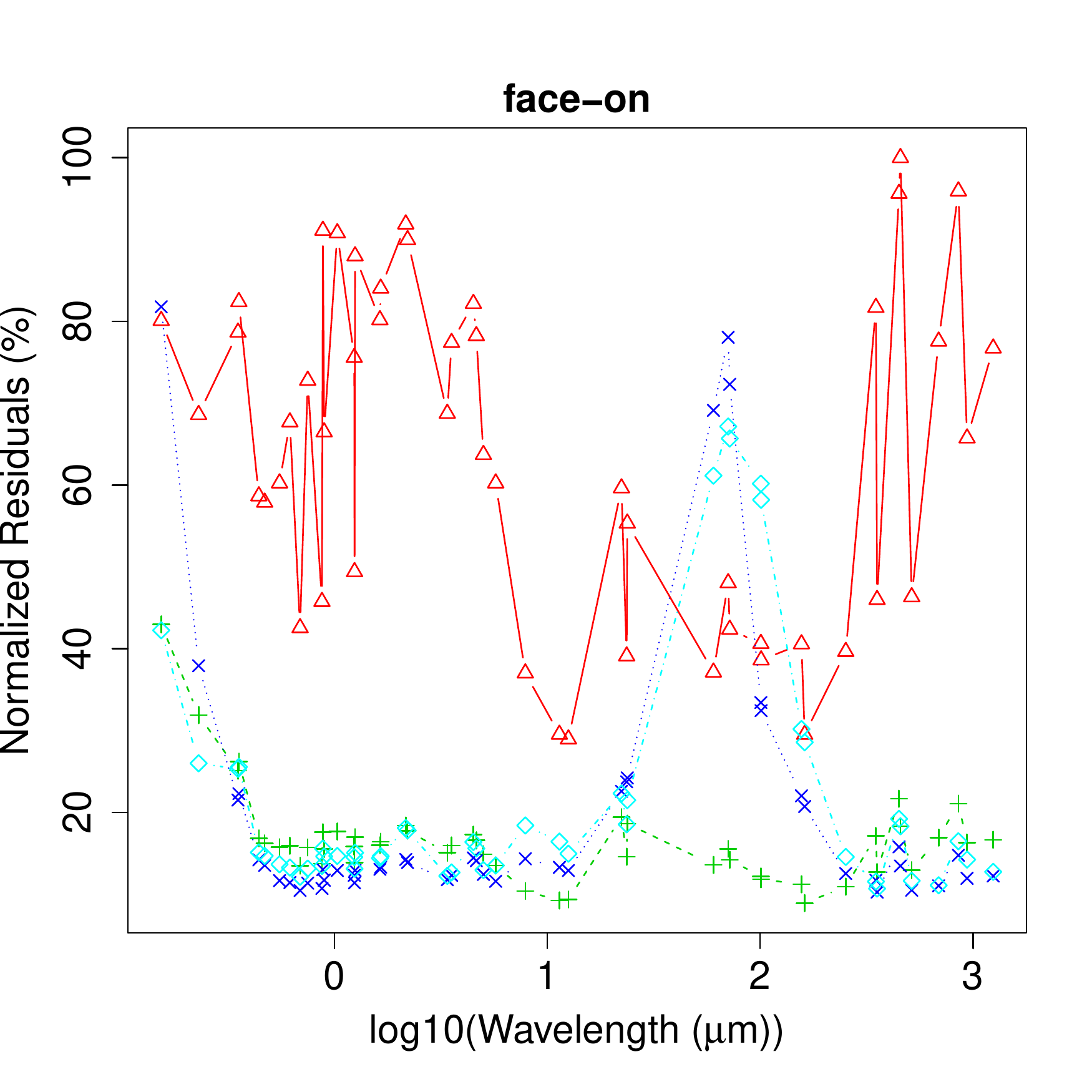}\hfill
\includegraphics[width=.5\textwidth]{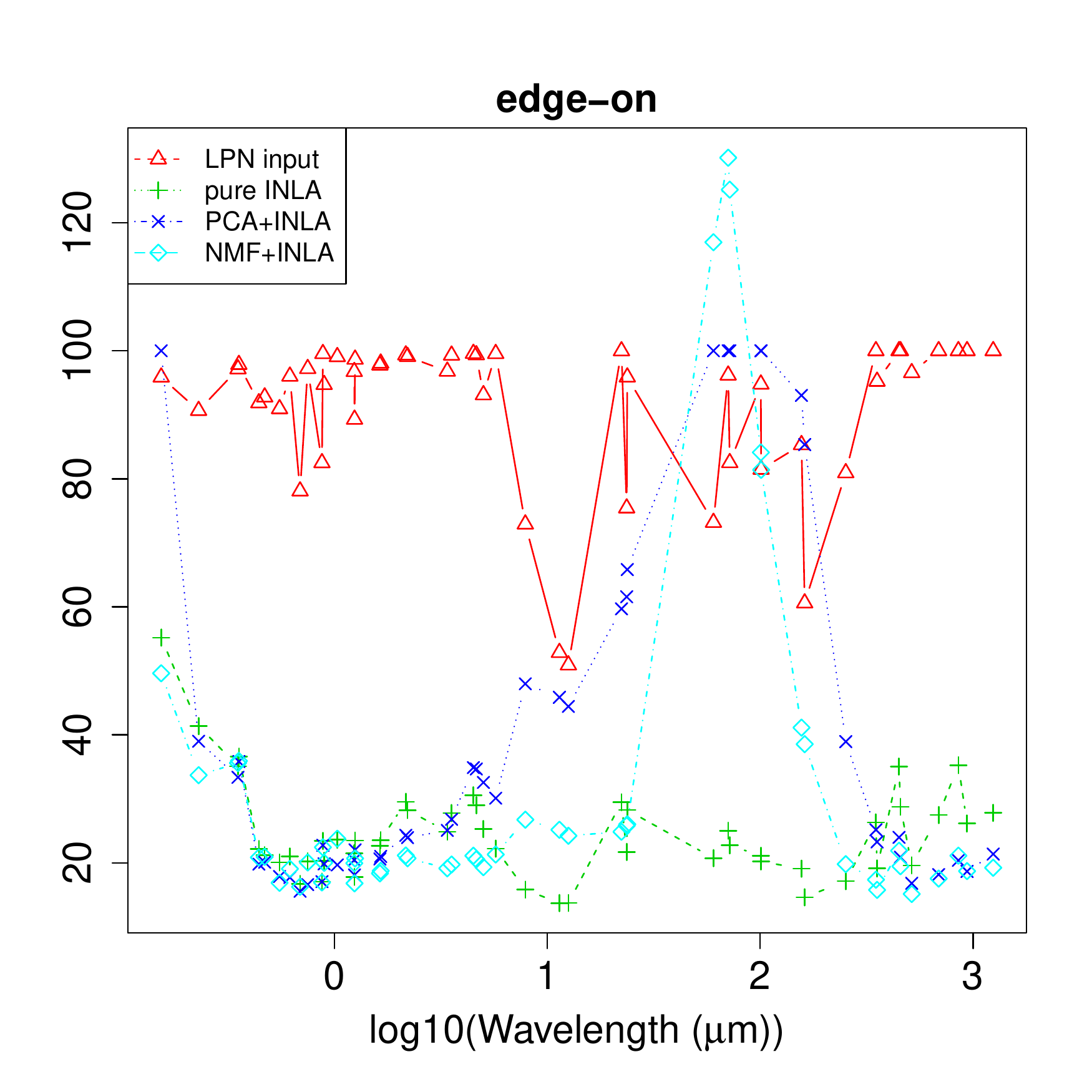}\hfill
\includegraphics[width=.5\textwidth]{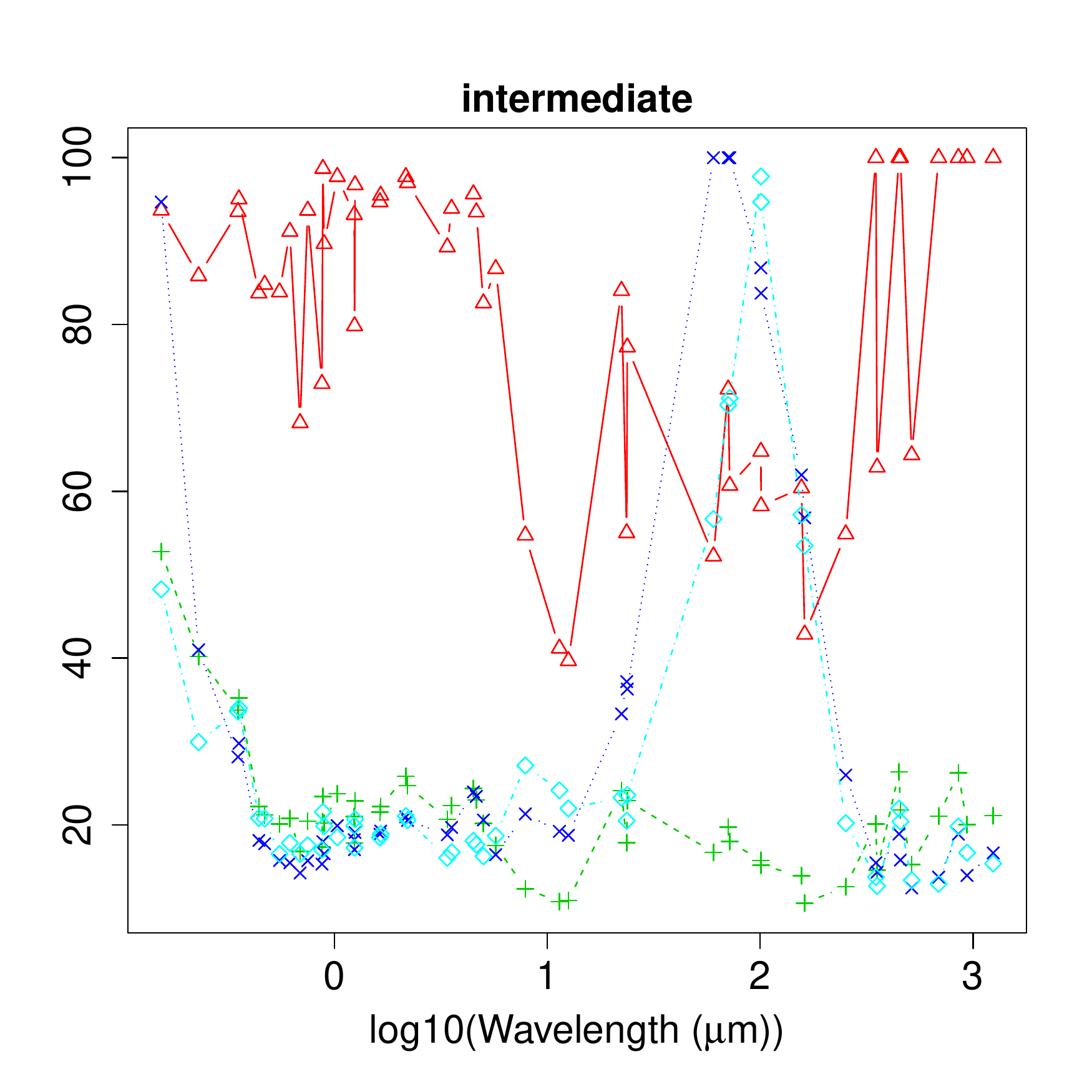}

}

\caption{Median of normalized residuals as a function of wavelength for 'face-on' (left), 'edge-on' (middle) and 'intermediate' (right) cubes.}
         \label{residuals_lambda}
   \end{figure*}

Figure \ref{residuals_lambda} shows the median of normalized residuals as a function of wavelength, for 'face-on' (left), 'edge-on' (middle) and 'intermediate' (right) cubes.
Although similar to  Figure \ref{integrated_sed} which shows normalized residuals for integrated SEDs, here we calculate 
residuals for each pixel at a given wavelength bin separately and then 
calculate their median value. 
Pixels with negative values are converted to zero, resulting in normalized residuals of 100\% (equation \ref{res_equatio}).
Overall, all of the employed methods (pure INLA and PCA/NMF+INLA)
results with significantly lower residuals compared to LPN input.
Pure INLA (green) is able to recover spatial information over the whole range of wavelength bins, with the median of normalized residuals $\lesssim20$ \%
for 'face-on' cube, and  $\lesssim30$ \% for 
'edge-on' and 'intermediate' cubes. 
On the other side, both PCA/NMF+INLA reconstructions have higher  residuals at wavelengths $\sim20-150~\mu$ m. 
At these wavelengths both LPN input and HPN reference images 
lose diffuse emission of low intensity that fills regions between spiral structure, otherwise present at other wavelength bins.
PCA and NMF components carry information about this diffuse emission and will always try to reconstruct it, resulting in higher residuals at wavelengths where the emission is 
attenuated in the original images.
However, statistics for these wavelength bins is governed by outskirt regions of the galaxy where PCA+INLA reconstructions fail, while NMF+INLA overestimates the flux density. 
When applied to a field cut that includes galaxy central region, both  PCA/NMF+INLA method results in lower residuals compared to pure INLA technique (Figure \ref{pca_res_cut}). Thus, masking out the outskirt regions prior INLA would improve the reconstructions.

\subsection{Summary and discussion}

\begin{table*}
\caption{Median of the normalized residuals and the running times for each realization, compared to HPN reference.}     
\centering
\label{results_table} \begin{tabular}{cclcllllll} 
\hline
                          &                                     & \multicolumn{1}{c}{LPN input} &                            & \multicolumn{2}{c}{pure INLA}                       & \multicolumn{2}{c}{PCA+INLA}                        & \multicolumn{2}{c}{NMF+INLA}                         \\
angle                     & photon                              & \multicolumn{1}{c}{median}    & input                      & median                   & time                     & median                   & time                     & median                   & time                      \\
                          & number                              & \multicolumn{1}{c}{(\%)}      & \multicolumn{1}{l}{sample} & \multicolumn{1}{c}{(\%)} & \multicolumn{1}{c}{(\%)} & \multicolumn{1}{c}{(\%)} & \multicolumn{1}{c}{(\%)} & \multicolumn{1}{c}{(\%)} & \multicolumn{1}{c}{(\%)}  \\ 
\hline
                          & $3\times10^9$                       & ~           ~17.68                       & 25\%                       & 9.97                     & 110                      & 10.68                    & 32                       & 11.35                    & 40                        \\
face-                     &                                     &                               & 10\%                       & 11.61                    & 89                       & 12.21                    & 26                       & 12.79                    & 35                        \\
on                        & $3\times10^8$                       & ~ ~60.97                       & 25\%                       & 15.57                    & 103                      & 15.46                    & 27                       & 17.65                    & 24                        \\
                          &                                     &                               & 10\%                       & 19.22                    & 74                       & 18.20                    & 15                       & 20.00                    & 19                        \\ 
\cline{2-10}
\multicolumn{1}{l}{}      & \multicolumn{1}{l}{$3\times10^{9}$} & ~ ~30.23                       & 25\%                       & 13.36                    & 58                       & 20.62                    & 18                       & 15.85                    & 24                        \\
\multicolumn{1}{l}{edge-} & \multicolumn{1}{l}{}                &                               & 10\%                       & 15.67                    & 43                       & 22.54                    & 15                       & 17.5                     & 20                        \\
on                        & \multicolumn{1}{l}{$3\times10^{8}$} & ~ ~96.88                       & 25\%                       & 23.42                    & 47                       & 31.18                    & 10                       & 24.92                    & 12                        \\
\multicolumn{1}{l}{}      & \multicolumn{1}{l}{}                &                               & 10\%                       & 29.38                    & 28                       & 35.28                    & 7                        & 27.47                    & 8                         \\ 
\cline{2-10}
\multicolumn{1}{l}{}      & $3\times10^9$                       & ~ ~24.91                       & 25\%                       & 12.10                    & 97                       & 12.63                    & 24                       & 12.73                    & 33                        \\
\multicolumn{1}{l}{inter-} & \multicolumn{1}{l}{}                &                               & 10\%                       & 14.57                    & 76                       & 14.41                    & 21                       & 14.51                    & 26                        \\
mediate                       & $3\times10^8$                       & ~ ~86.13                       & 25\%                       & 20.03                    & 89                       & 19.27                    & 22                       & 23.00                    & 22                        \\
\multicolumn{1}{l}{}      & \multicolumn{1}{l}{}                &                               & 10\%                       & 25.10                    & 66                       & 22.63                    & 13                       & 25.89                    & 16                        \\
\hline
\end{tabular}
\end{table*}

Table \ref{results_table} summarises statistics for the spatial reconstructions of 'face-on', 'edge-on' and 'intermediate' cubes, using pure INLA
and PCA/NMF+INLA implementations. We present results for two different LPN input images ($3\times10^{8}$ and $3\times10^{9}$ photon number), using randomly sampled 10\% or 25\% of the available spatial information.
Both LPN input images and the reconstructions are compared to HPN reference. 
The median of the normalized residuals are shown for each cube, together with the total running time required for LPN \textsc{skirt} simulations and INLA reconstructions. Running times presented in Table \ref{results_table} is the following:

\begin{align}
 t~(\%) = \frac{t_{\rm{LPN~\textsc{skirt}}}+t_{\rm{INLA}}}{t_{\rm{HPN~\textsc{skirt}}}} *100,
\end{align}

\noindent where $t_{\rm{INLA}}$ refers either to pure INLA reconstructions
(50 images at 50 wavelength bins) or to the total time for PCA/NMF analysis and PCA/NMF+INLA reconstructions.

For each realization, residuals for our reconstructions are significantly lower compared to LPN input residuals, proving that each of the employed methods is able to recover spatial structure using only 10\% or 25\% of the LPN input. 
Although pure INLA reconstructions result in the lowest residuals,
this methodology
does not provide a significant speedup. Using  a
sample size of 10\% requires $\gtrsim70$\% of HPN reference \textsc{skirt} running time while using 25\% of 'face-on' cube exceeds it. 
However, employing dimensionality reduction techniques successfully
reduces the running times up to $\sim$ 10\%  of HPN reference's
with similar residuals.

Additionally, the quality of our reconstructions scales with LPN input residuals, thus the 'face-on' cube has overall lower residuals compared
to other tilt angles.
Depending on the LPN input ($3\times10^{8}$ or $3\times10^{9}$ photon number), and the sample size (10\% or 25\%) residuals of PCA/NMF+INLA reconstructions are in range $\sim$ 10-20\% for 'face-on' cube, and $\sim$ 15-30\% for
'edge-on' and 'intermediate' cubes. 
Naturally, the lowest residuals are achieved when using 25\% of 
$3\times10^{9}$ photon number LPN input cube, which requires 
$\sim$ 25-40\% of the HPN reference running time.
At the other end, by sampling 10\% of the $3\times10^{8}$ photon number LPN input cube, the
running times are reduced to $\sim$7-20\% of the HPN reference.
The 'edge-on' cube typically has shorter running times due to a smaller size field (2401$\times$901$\times$50), compared to the 'face-on' cube (2151$\times$2151$\times$50).

When applied to a field cut, PCA+INLA method outperforms 
pure INLA, both by lower residuals and significantly shorter
running times (Figure \ref{pca_res_cut}),
since this cut does not include galaxy outskirts. This proves that performing a careful prior segmentation of the galaxy or astrophysical source will aid best in our post-processing technique.
Furthermore, PCA analysis has advantages over NMF:
PCA analysis is faster and the number of PCA components does not need to be 
decided in advance. On the other hand, NMF provides always physically positive reconstructions, particularly useful at low flux regions like the outskirts of the image.

\section{Conclusions}
\label{conclusions}

In this work, we use PCA and NMF dimensionality reduction techniques 
together with approximate Bayesian inference of continuous Gaussian random fields with INLA for \textsc{skirt} simulations post-processing. We test our methodology using three images of Au-16 \textsc{skirt} Auriga galaxy, simulated with different tilt angles: 'face-on', 'edge-on', and 'intermediate'.
These HPN reference images ($3\times10^{10}$ photon packages) served as the 'ground truth' to which
we compared the performance of our method applied to LPN input images ( $3\times10^{8}$ or  $3\times10^{9}$ photon packages).

Our results showed that spatially integrated SED closely follow the reference SED for each of the employed methods, with 
the median of the normalized residuals typically $\lesssim0.3\%$.
Spatial modelling suggests that the quality of our reconstructions
changes with the position along the galaxy plane, with
faint galaxy outskirts having the highest residuals. 

Depending on the number of photon packages, the sample size
and the desired quality of reconstruction, our method offers time-efficient
reconstructions with spatial residuals $\sim20-30\%$, requiring 
$\sim7-20\%$ of the HPN reference running time. 
Higher quality reconstructions can be achieved by sampling 25\% of $3\times10^{9}$ LPN input image, resulting in residuals
$\sim10-20\%$ within $\sim20-40\%$ of the HPN reference running time.

In order to improve the quality of reconstructions, different sampling methods based on information on the spatial densities, and different pre-INLA data pre-processing might be worth exploring.
Even though our results are not sensitive to the choice of sampling method (random or uniform), more complex and less homogeneous spatial maps might require non-uniform sampling.

Being able to efficiently perform large amounts of numerical simulations with varying physical characteristics is essential to compare to real observations in a quantitative way. Even if the accuracy is not immediately as good as that of a HPN simulation, such explorations with optimized LPN  simulations can help narrowing down the parameters to then run a full HPN simulation. This work represents a crucial step in this direction.

\begin{acknowledgements}
We acknowledge the support by the programme of scientific and technological cooperation between the government of the Republic of Serbia and the government of the Republic of Portugal, (grant~No.~337-00-00227/2019-09/53 and FCT 5581 DRI, Sérvia 2020/21). M.S. and M.S. acknowledge support by the Science Fund of the Republic of Serbia, PROMIS 6060916, BOWIE and by the Ministry of Education, Science and Technological Development of the Republic of Serbia through the contract no. 451-03-9/2022-14/200002.
J. R.-S. is funded by Funda\c{c}\~ao para a Ci\^encia e a Tecnologia (PD/BD/150487/2019), via the International Doctorate Network in Particle Physics, Astrophysics and Cosmology. S.G.G and J.R.-S. acknowledge the support of FCT under Project CRISP PTDC/FIS-AST-31546/2017.
\end{acknowledgements}

%
%

\bibliographystyle{aa} 
\bibliography{references} 

\begin{appendix} 

\section{Spatial reconstructions of PCA/NMF components}
\label{color_maps_components}

\begin{figure*}[h]
\captionsetup[subfigure]{labelformat=empty}

 \centering
    \subfloat[]{\includegraphics[width=0.37\textwidth]{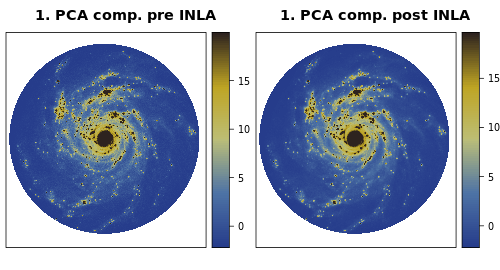}}
    \subfloat[]{\includegraphics[width=0.37\textwidth]{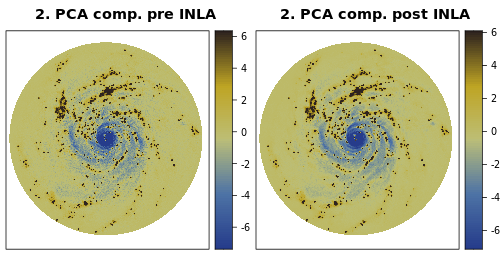}}

    \vfill
    \subfloat[]{\includegraphics[width=0.37\textwidth]{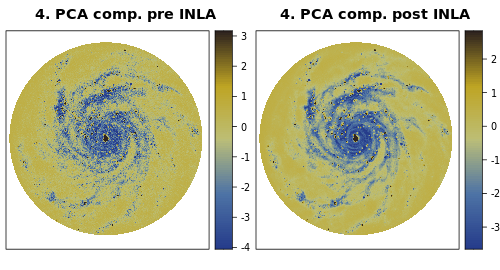}}
    \subfloat[]{\includegraphics[width=0.37\textwidth]{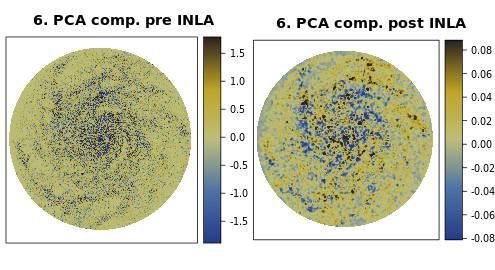}}

    \comment{
    \vfill
    \subfloat[]{\includegraphics[width=0.37\textwidth]{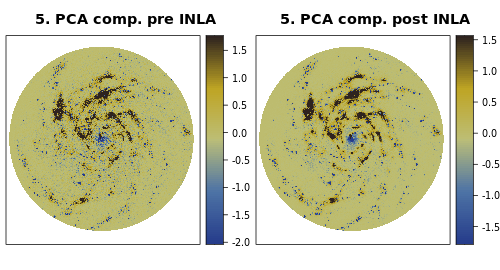}}
    \subfloat[]{\includegraphics[width=0.37\textwidth]{color_map6_pca.png}}

    \vfill
    \subfloat[]{\includegraphics[width=0.37\textwidth]{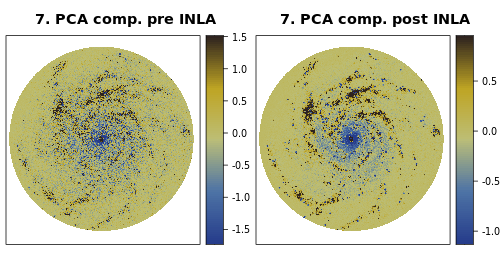}}
    \subfloat[]{\includegraphics[width=0.37\textwidth]{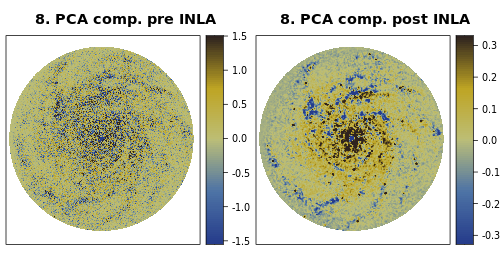}}

    \vfill   
    \subfloat[]{\includegraphics[width=0.37\textwidth]{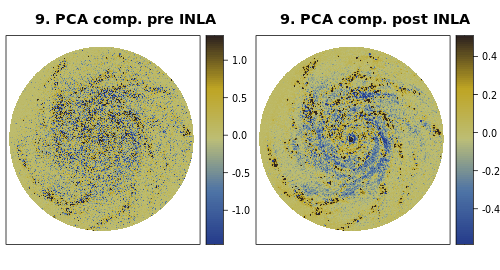}}
    \subfloat[]{\includegraphics[width=0.37\textwidth]{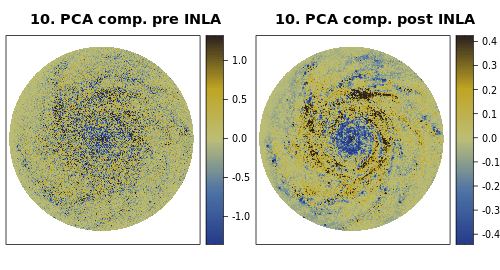}}
    
}
\caption{Spatial maps of PCA coefficients
attached to principal components before INLA (left panels)
and after INLA (right panels). Sample size: 25\%, 'face-on' cube.
}
         \label{color_maps_pca}
   \end{figure*}

\begin{figure*}[h]
\captionsetup[subfigure]{labelformat=empty}

 \centering
    \subfloat[]{\includegraphics[width=0.37\textwidth]{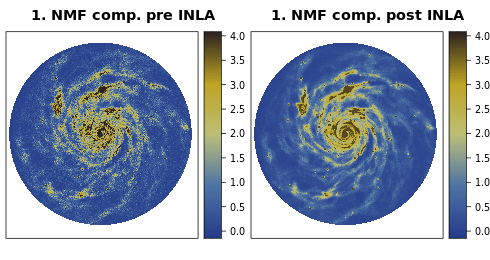}}
    \subfloat[]{\includegraphics[width=0.37\textwidth]{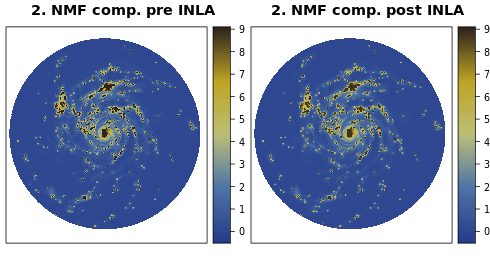}}

    \vfill
    \subfloat[]{\includegraphics[width=0.37\textwidth]{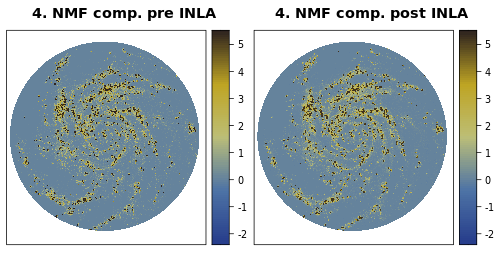}}
    \subfloat[]{\includegraphics[width=0.37\textwidth]{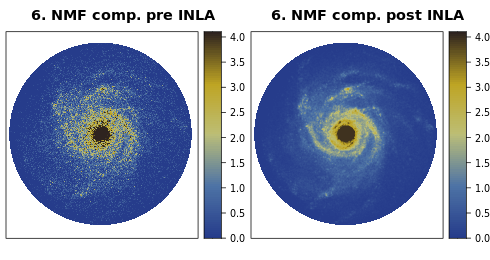}}
 \comment{
    \vfill
    \subfloat[]{\includegraphics[width=0.37\textwidth]{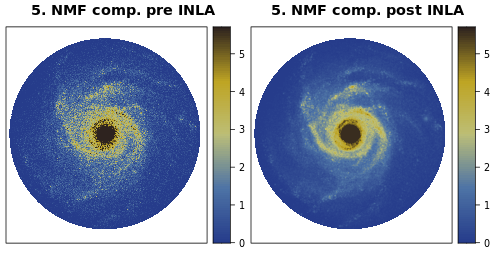}}
    \subfloat[]{\includegraphics[width=0.37\textwidth]{color_map6_nmf.png}}

    \vfill
    \subfloat[]{\includegraphics[width=0.37\textwidth]{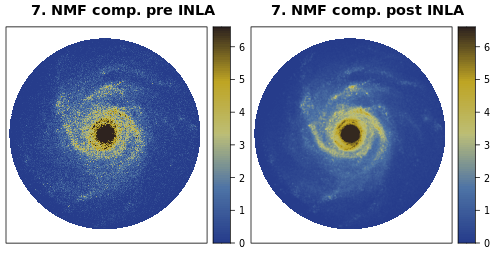}}
    \subfloat[]{\includegraphics[width=0.37\textwidth]{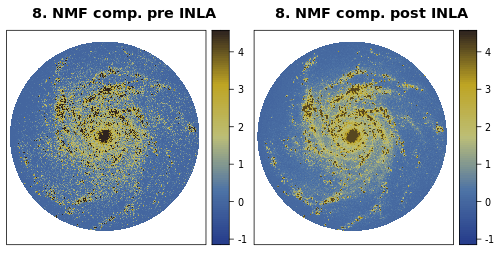}}

    \vfill   
    \subfloat[]{\includegraphics[width=0.37\textwidth]{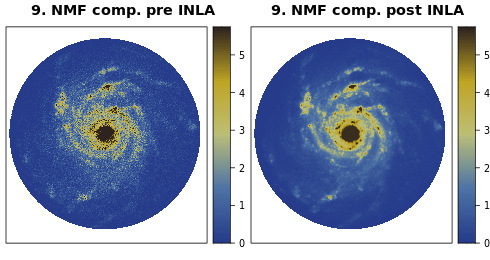}}
    \subfloat[]{\includegraphics[width=0.37\textwidth]{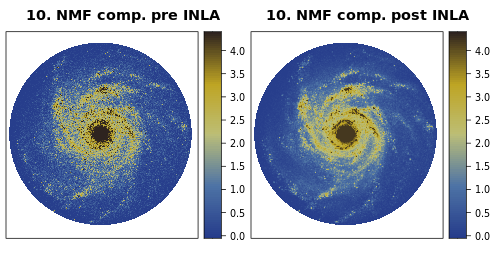}}
    }

\caption{Same as Fig. \ref{color_maps_pca}, for NMF method.
}
         \label{color_maps_nmf}
   \end{figure*}

Here, we further explore the 
reconstructions obtained using PCA and NMF methods.
Figures \ref{color_maps_pca} and \ref{color_maps_nmf}
show examples of spatial maps of PCA and NMF coefficients
attached to principal components before INLA (left panels)
and after INLA (right panels). 
PCA sorts principal components by the amount of information they carry, which can be seen in Figure \ref{color_maps_pca}. 
The first component reconstructs the highest flux density regions:
galaxy centre and strong spiral structure.
The second principal component reconstructs outer parts of the spiral arms and the galaxy outskirts.
The remaining principal components reconstruct background emission and less prominent features in the spiral structure.
Up to the sixth principal component, INLA reconstructions 
do not significantly differ from the input. 
However, for the remaining principal components INLA reconstructions predict less variability, which is especially evident for the sixth 
component.

On the other hand, NMF components are not sorted by the amount of variability. Figure \ref{color_maps_nmf} shows that NMF components can be distinguished between those that reconstruct central parts, spiral structure, and a mixture of different features. Overall, the structure
seen in all NMF components is preserved and slightly highlighted 
after INLA reconstructions.

\clearpage

\clearpage

   \section{Pure PCA/NMF reconstructions}
\label{pure_reconstructions}
   
\begin{figure*}[h]
\captionsetup[subfigure]{labelformat=empty}

 \centering
    \subfloat[0.23 $\mu$m]{\includegraphics[width=0.4\textwidth]{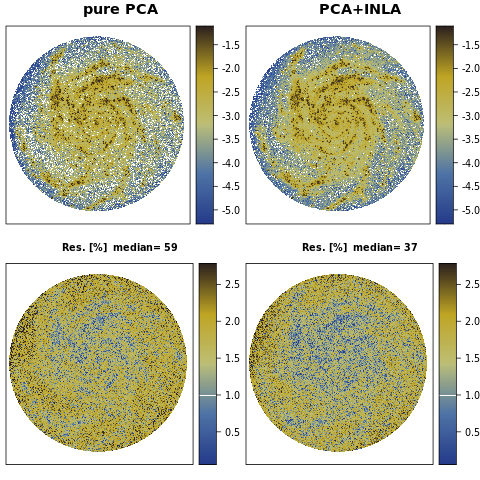}} 
    \subfloat[7.88 $\mu$m]{\includegraphics[width=0.4\textwidth]{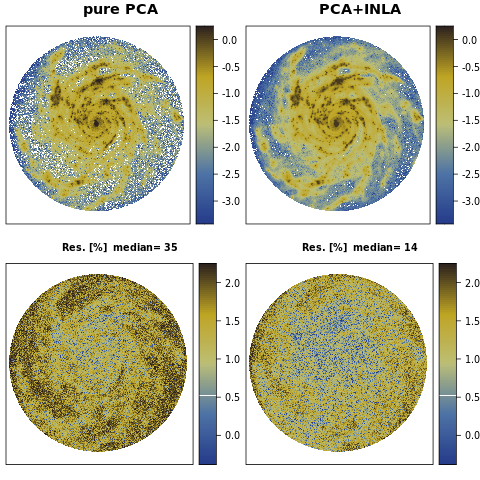}} 

    \vfill
    \subfloat[100.08 $\mu$m]{\includegraphics[width=0.4\textwidth]{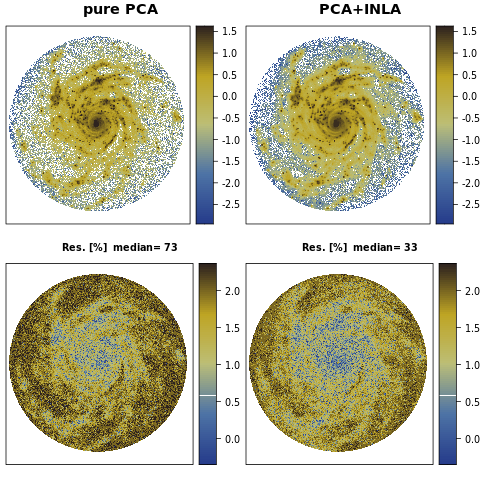}}
    \subfloat[515.36 $\mu$m]{\includegraphics[width=0.4\textwidth]{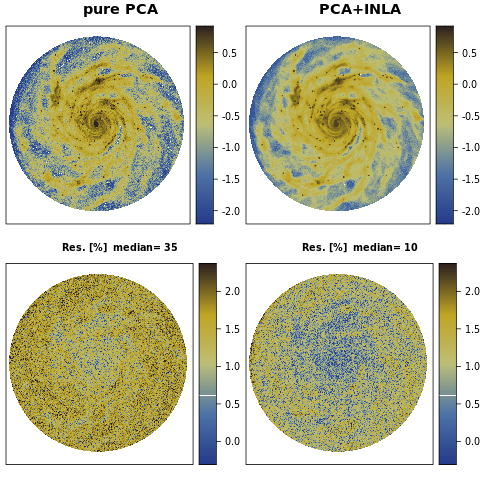}}

\caption{Pure PCA and PCA+INLA spatial reconstructions (upper panels)
and the associated residuals (lower panels)
at wavelength bins:
0.23 $\mu$m, 7.88 $\mu$m, 100.08 $\mu$m and 515.36 $\mu$m. 
Sample size: 25\%, 'face-on' cube.
}
         \label{color_maps_pure_pca}
         
   \end{figure*}

\begin{figure*}[h]
 \centering
 \captionsetup[subfigure]{labelformat=empty}

    \subfloat[0.23 $\mu$m]{\includegraphics[width=0.4\textwidth]{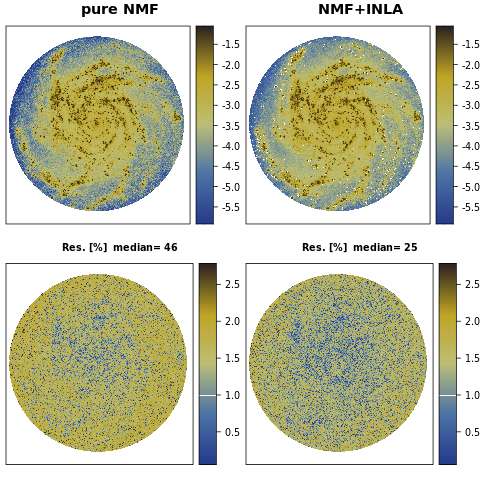}} 
    \subfloat[7.88 $\mu$m]{\includegraphics[width=0.4\textwidth]{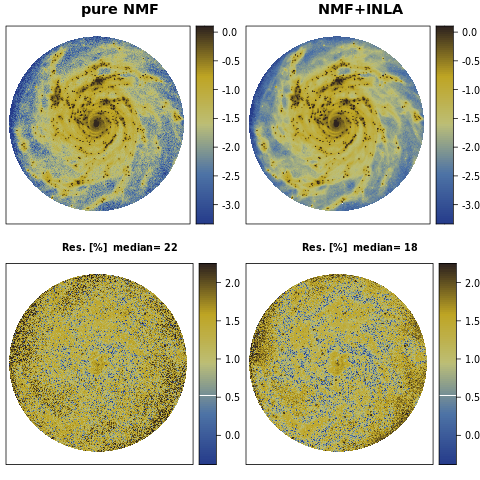}} 

    \vfill
    \subfloat[100.08 $\mu$m]{\includegraphics[width=0.4\textwidth]{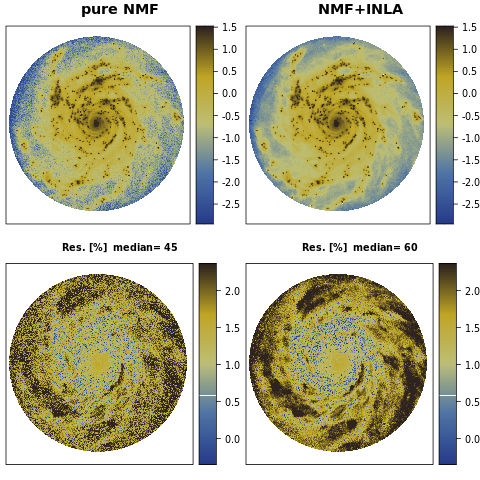}}
    \subfloat[515.36 $\mu$m]{\includegraphics[width=0.4\textwidth]{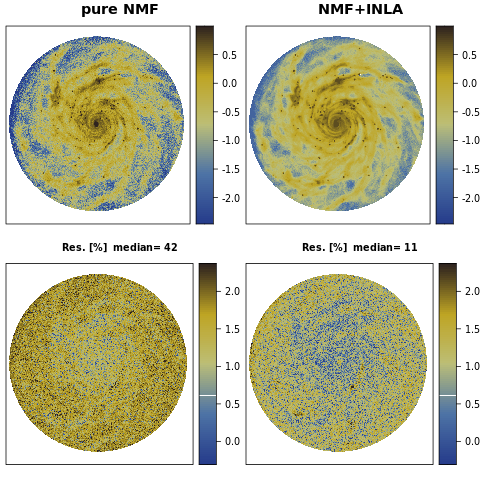}}

\caption{Same as Fig. \ref{color_maps_pure_pca} for pure NMF and NMF+INLA reconstructions.
}
         \label{color_maps_pure_nmf}
         
   \end{figure*}

In this section we compare PCA (Figure \ref{color_maps_pure_pca})
and NMF (Figure \ref{color_maps_pure_nmf}) reconstructions before and after INLA. 
Figures \ref{color_maps_pure_pca} and \ref{color_maps_pure_nmf}
show pure PCA/NMF and PCA/NMF+INLA reconstructions 
(upper panels) and the associated residuals (lower panels)
at different wavelength bins:
0.23 $\mu$m, 7.88 $\mu$m, 100.08 $\mu$m and 515.36 $\mu$m.
Pure PCA/NMF reconstructions refer to 
results of PCA or NMF analyses performed using 100\% of full LPN input cube and ten PCA/NMF components.
The quality of these reconstructions is further enhanced with 
INLA using  randomly sampled 25\% of PCA/NMF coefficients.
Results of both pure PCA/NMF and PCA/NMF+INLA reconstructions 
are compared to HPN reference images.

Figure \ref{color_maps_pure_pca} shows that pure PCA method produces pixels with negative values. PCA/NMF+INLA reconstructions result in lower residuals compared to the pure PCA/NMF technique.
Employing INLA  reduces the amount of negative reconstructions leading to lower residuals and thus better quality of reconstructions. 
However, at wavelengths around 
$\sim100$ $\mu$m the amount of negative values is still high after INLA, leading to higher residuals at the galaxy outskirts.
On the other hand, NMF reconstructions (Figure \ref{color_maps_pure_nmf}) do not suffer from nonphysically negative reconstructions, but at the same wavelength bins ($\sim100$ $\mu$m ) NMF+INLA overestimate flux densities
at the galaxy outskirts, leading to higher residuals compared to pure NMF reconstructions. 

\end{appendix}

\end{document}